\documentclass[10pt,journal,compsoc]{IEEEtran}
\usepackage{color,soul}
\usepackage{comment}
\usepackage{cite}
\usepackage{graphicx}
\usepackage{hyperref}
\usepackage[table,xcdraw]{xcolor}
\usepackage{multirow}
\usepackage{pifont}
\usepackage{stix}
\usepackage{tikz}
\usepackage{microtype}
\usepackage{setspace}
\newcommand*\circled[1]{\tikz[baseline=(char.base)]{
            \node[shape=circle,fill=black,text=white,draw,inner sep=0.3pt] (char) {\footnotesize #1};}} 

\begin{document}

\sloppy

\title{Maximal Extractable Value in Decentralized Finance: Taxonomy, Detection, and Mitigation}

\author{Huned Materwala, 
Shraddha M. Naik, 
Aya Taha{*},
Tala Abdulrahman Abed{*}, and
Davor~Svetinovic
\thanks{All authors are with the Center for Secure Cyber-Physical Systems, Department of Computer Science, Khalifa University, Abu Dhabi, UAE.\\{*}These authors contributed equally to this work.\\ D. Svetinovic is the corresponding author: davor.svetinovic@ku.ac.ae.
}}

\markboth{Journal of \LaTeX\ Class Files,~Vol.~xx, No.~x, October~2024}%
{Shell \MakeLowercase{\textit{et al.}}: A Sample Article Using IEEEtran.cls for IEEE Journals}

\maketitle

\begin{abstract}
Decentralized Finance (DeFi) leverages blockchain-enabled smart contracts to deliver automated and trustless financial services without the need for intermediaries. However, the public visibility of financial transactions on the blockchain can be exploited, as participants can reorder, insert, or remove transactions to extract value, often at the expense of others. This extracted value is known as the Maximal Extractable Value (MEV). MEV causes financial losses and consensus instability, disrupting the security, efficiency, and decentralization goals of the DeFi ecosystem. Therefore, it is crucial to analyze, detect, and mitigate MEV to safeguard DeFi. Our comprehensive survey offers a holistic view of the MEV landscape in the DeFi ecosystem. We present an in-depth understanding of MEV through a novel taxonomy of MEV transactions supported by real transaction examples. We perform a critical comparative analysis of various MEV detection approaches, evaluating their effectiveness in identifying different transaction types. Furthermore, we assess different categories of MEV mitigation strategies and discuss their limitations. We identify the challenges of current mitigation and detection approaches and discuss potential solutions. This survey provides valuable insights for researchers, developers, stakeholders, and policymakers, helping to curb and democratize MEV for a more secure and efficient DeFi ecosystem.
\end{abstract}


\section{Introduction}
\IEEEPARstart{D}{ecentralized Finance} (DeFi) \cite{puschmann2024taxonomy}, powered by blockchain technology, offers financial services, such as investing, lending loans, and trading assets, to various stakeholders without the need for intermediary brokers. This creates a decentralized, secure, transparent, and traceable financial ecosystem. According to a Skyquest report, the DeFi market size is expected to reach 48.02 billion USD by 2031, up from 23.99 billion USD in 2023, a compound annual growth rate of 9.06\%\footnote{\url{https://www.skyquestt.com/report/decentralized-finance-market}, accessed on 19 September 2024}. Furthermore, as of September 2022, the total value locked in the DeFi ecosystem exceeded 82 billion USD\footnote{\url{https://defillama.com/}, accessed on 19 September 2024}, with the Ethereum network accounting more than 45 billion USD (approximately 56\%).

Decentralized Exchanges (DEXes) are one of the prominent applications of DeFi, enabling users to directly swap tokens using smart contracts \cite{xu2023sok}. Most DEXes operate on the Ethereum blockchain to facilitate trustless and automated transactions. However, in the Ethereum network, transactions are stored in a public mempool before they are included in a block. This creates profitable opportunities for network participants, referred to as searchers, who can submit new transactions by observing pending financial transactions to gain additional revenue. These searchers maximize their profits by manipulating gas prices to influence the order of their transactions within the block. This additional value extracted from the blockchain network is termed Maximal Extractable Value (MEV) {\cite{babel2023clockwork}. MEV searchers can be the block producers (i.e., miners, validators, or block proposers), other network participants, or bots. However, block producers have a unique advantage, as they can include transactions in blocks without paying high gas prices \cite{torres2021frontrunner}. Furthermore, the transition from Proof-of-Work (PoW) to Proof-of-Stake (PoS) in Ethereum has significantly reduced block producer rewards, further luring block producers to engage in MEV activities \cite{kapengut2023event}.

MEV could result in major financial losses, network congestion, increased gas prices, and blockchain inefficiency \cite{torres2021frontrunner,poux2022maximal}. Before Ethereum's transition to PoS in September 2022, around 440,000 ETH in MEV was extracted. Since the transition, approximately 180,000 ETH has been extracted up until May 2023\footnote{\url{https://milkroad.com/guide/mev/}, accessed on 19 September 2024}. However, not all MEV transactions are detrimental to the DeFi ecosystem. Some can destabilize consensus mechanisms and cause economic harm, while others may even stabilize financial markets \cite{mohan2024blockchains}. Therefore, understanding, detecting, and mitigating MEV is critical to ensure the efficiency and stability of the DeFi ecosystem.

Several surveys have explored various aspects of MEV, including transaction types and mitigation strategies \cite{li2022survey,alipanahloo2024maximal,mohan2024blockchains,poux2022maximal,rasheed2024mevecosystem,gramlich2024maximal}. In particular, \cite{li2022survey} classifies MEV transactions, while \cite{alipanahloo2024maximal,mohan2024blockchains} emphasize on mitigation techniques. In contrast, \cite{poux2022maximal,rasheed2024mevecosystem,gramlich2024maximal} cover both transaction types and mitigation strategies. However, none have thoroughly examined all transaction types, detection approaches, and mitigation strategies. This comprehensive survey addresses this gap by presenting a novel taxonomy of MEV transaction types supported by real-world examples from the Ethereum network. It also explores MEV detection approaches and mitigation strategies and examines various simulation and extraction methods. The main contributions of this survey are as follows.

\begin{itemize}
    \item We introduce a novel and comprehensive taxonomy of MEV transactions, supported by real Ethereum transaction examples for each identified MEV type. The taxonomy distinguishes between value-diverting MEV transactions, which can lead to financial loss and network instability, and value-creating MEV transactions, which can stabilize markets or enhance efficiency. This distinction provides clarity on the dual nature of MEV activities and lays the foundation for discussions on detection and mitigation strategies.

    \item We present a critical comparative analysis of the various MEV detection approaches proposed for different types of MEV transactions, highlighting their effectiveness in identifying these transaction types.

    \item We provide an in-depth analysis of various MEV mitigation strategies, such as transaction ordering solutions, privacy-preserving public pools, and private pools. By identifying the specific limitations of these strategies, we offer a critical assessment of their real-world applicability and potential areas for improvement.

    \item We explore existing MEV simulation frameworks and extraction methods, offering a comprehensive comparison of how these methods model and replicate real-world MEV scenarios.

    \item We critically examine the major challenges in MEV detection, mitigation, and simulation, such as centralization, latency, cross-domain MEV, multi-address MEV, and simulation environment. Furthermore, we discuss potential solutions paving the way for a more secure and efficient DeFi system.
\end{itemize}

The remainder of this article is organized as follows. Section 2 provides essential background on Ethereum, DEXes, and MEV, enabling readers to gain a clearer understanding of the key concepts. Section 3 reviews existing related surveys on MEV, highlighting gaps in the literature. In Section 4, we describe the methodology used to conduct this survey. Section 5 introduces and synthesizes the proposed taxonomy of MEV transactions. Sections 6 and 7 focus on MEV detection approaches and mitigation strategies, respectively. Section 8 explores MEV simulation and extraction methods. Section 9 addresses the challenges associated with MEV mitigation and detection, along with potential solutions. Finally, Section 10 concludes the survey.

\section{Background}
This section explains the concepts and terminology related to Ethereum, DEXes, and MEV to facilitate a better understanding of the remainder of the paper.

\subsection{Ethereum}
Transaction execution in Ethereum requires gas due to the computational work involved \cite{zulfiqar2021ethreview}. Users set a gas price for each transaction they submit, which represents the amount they are willing to pay per unit of gas consumed. The total transaction fee, known as the gas fee, is calculated by multiplying the gas price by the amount of gas used \cite{daian2020flash}. Each transaction must also specify a gas limit, which determines the maximum amount of gas the user is willing to allocate for the transaction's execution. The gas limit helps ensure that the user has sufficient funds to cover the gas fee by verifying that the available balance exceeds the product of the gas price and the gas limit. Additionally, the gas limit prevents transactions from running indefinitely, as the transaction will be aborted if gas consumption reaches the specified limit \cite{daian2020flash}.

The transactions are mined into a block by the block producer (miner in case of PoW and validator in case of PoS). The gas fee for each transaction in the block is then paid to the block producer as a reward. However, the number of transactions in a block is restricted by the block gas limit, which determines the maximum gas consumption of all the transactions combined in that block. Consequently, to increase the reward, the block producers prioritize the transactions with higher gas prices. This leads to instability as the users would pay high gas prices to ensure the inclusion of their transactions. To address this issue, Ethereum Improvement Proposal-1599 (EIP-1599) \cite{roughgarden2020transaction} was proposed and made effective in August 2021.

\subsubsection{EIP-1599}
With the introduction of EIP-1599, transaction gas fees are now divided into a base fee and a priority fee \cite{mohan2024blockchains}. The base fee is the minimum amount required for a transaction to be included in a block. Instead of being paid as a reward to the block producer, the base fee is burned (i.e., sent to an inaccessible address) upon transaction execution. The priority fee, however, is paid directly to block producers to incentivize them to include the transaction in a block. In addition to the priority fee, producers may also receive direct private transfers from users as additional incentives \cite{sokolov2021ransomware}.

\subsection{Decentralized Finance (DeFi) Applications}

DeFi applications offer a wide range of financial services through decentralized protocols \cite{aquilina2024decentralized,gramlich2024maximal}. Below, we provide an overview of the widely used DeFi applications. This will help in understanding how various MEV transactions can impact these applications.

\textit{DEXes} are blockchain-based trading platforms that eliminate the need for a centralized operator. Unlike traditional centralized exchanges (CEXes), where an exchange operator maintains an order book and facilitates trade execution \cite{makarov2020trading}, DEXes use smart contracts to enable direct peer-to-peer asset trading \cite{xu2023sok}. The trading mechanisms in DEXes can be either order book-based, where orders are matched manually (as in EtherDelta) or automatically by a smart contract (as in IDEX) \cite{daian2020flash}, or based on Automated Market Makers (AMMs), which use smart contracts to create liquidity pools. In AMMs, users trade assets directly against the pool, with prices determined algorithmically based on the new ratio of assets in the pool. Xu et al. \cite{xu2023sok} provides a detailed analysis of AMM pricing models. Larger asset trades result in greater price shifts, leading to slippage, where traders receive a less favorable price than expected due to the impact of their trade on the pool’s balance.

\textit{Lending and borrowing} platforms in DeFi allow users to lend and borrow assets from lending pools. Lenders earn interest paid by borrowers, who must provide collateral as security. The value of the collateral must exceed a minimum ratio, which varies depending on the lending protocol and asset (e.g., 150\% for certain assets). If the collateral's value falls below the liquidation ratio due to market volatility, liquidators can repay part of the loan in exchange for acquiring the collateral at a discounted price. This process, known as liquidation, ensures that lenders are repaid and prevents bad debt accumulation. Borrowers can avoid liquidation by depositing additional collateral or repaying part of the loan to maintain a safe ratio.

\textit{Yield farming} enables users to generate returns by depositing assets into liquidity pools or lending protocols in exchange for interest and trading fees. On the other hand, \textit{Margin trading} allows users to borrow funds to leverage their trades, with smart contracts automating the borrowing, lending, and liquidation processes. \textit{Derivatives} in DeFi enable users to trade financial contracts (derivatives) that are linked to underlying crypto assets. \textit{Asset management} platforms allow users to optimize and manage their cryptocurrency holdings using tools for portfolio management, risk management, and yield optimization. In \textit{prediction markets}, users can place bets on the outcomes of future events, such as elections, sports matches, or financial market movements, with decentralized oracles ensuring trustless and transparent verification.

\textit{Tokenization} in DeFi has made it possible to convert real-world physical assets like real estate and commodities into digital tokens, enabling their trade within the DeFi ecosystems. \textit{Domain name protocols} allows the creation of censorship-resistant decentralized domains to host websites or services that are not subject to traditional censorship or regulation. To enhance the security of DeFi protocols, some platforms reward users for identifying \textit{vulnerabilities in smart contracts}. Lastly, \textit{Gambling and lottery} platforms within DeFi offer transparent and trustless participation gambles and lotteries, with outcomes governed by smart contracts.

\subsection{Maximal Extractable Value (MEV)}
Publicly visible transactions on Ethereum enable searchers to extract MEV through the insertion, removal, or reordering of transactions \cite{daian2020flash}. According to Ethereum's documentation\footnote{\url{https://ethereum.org/en/developers/docs/mev/}, accessed on 10 February 2024}: \textit{Maximal extractable value (MEV) refers to the maximum value that can be extracted from block production in excess of the standard block reward and gas fees by including, excluding, and changing the order of transactions in a block.}

Before the Merge\footnote{The update that transitioned Ethereum's consensus protocol from Proof-of-Work (PoW) to Proof-of-Stake (PoS) to improve scalability, security, and energy efficiency}, Ethereum was based on the PoW consensus protocol, where block producers were known as miners. The value extracted from block production was thus initially referred to as Miner Extractable Value. After the Merge in September 2022, Ethereum transitioned to the PoS consensus protocol, with validators taking the place of miners. Additionally, a searcher does not necessarily have to be the block producer; they can also be non-miner users or bots. Consequently, the term "Miner" in MEV was replaced with "Maximal" to reflect this change. MEV is sometimes also referred to as "Blockchain Extractable Value" \cite{qin2022quantifying}. Furthermore, the Flashbots project refers to the actual extracted values due to MEV opportunities as "Realized Extractable Value"\footnote{\url{https://writings.flashbots.net/quantifying-rev}, accessed on 05 September 2024}. Judmayer et al. \cite{judmayer2022estimating} describe the potential or future values that could be extracted if a sequence of transactions eventually becomes a part of the blockchain (with some probability) as "Expected Extracted Value".

\subsubsection{MEV Transactions}

The main aspects of MEV include the actions performed to extract MEV (i.e., insertion, removal, or reordering), the network participants who act as searchers, and the nature of MEV transactions (i.e., single or multi-block). Based on these aspects, we define MEV transactions as \textit{transactions inserted by searchers (block producers, bots, or other network participants) to gain a financial advantage in DeFi by exploiting victim transactions\footnote{A transaction in the public mempool that presents an opportunity for a searcher to generate profits.} or market fluctuations. These transactions can span multiple blocks, with their order determined through gas price adjustments and/or private tips.} Searchers can temporarily censor a victim transaction by executing multiple MEV transactions. It is important to note that not all MEV transactions are inherently harmful. Some enhance market efficiency, while others lead to financial losses, network congestion, or blockchain instability.

For instance, a searcher may identify a price difference for the same asset across different DeFi protocols due to trading activities. The searcher performs an MEV transaction to profit from this price difference, which helps stabilize asset prices across these platforms. Such transactions are known as "value-creating" MEV transactions, as they generate new value \cite{mohan2024blockchains}. These transactions are generally regarded as beneficial rather than harmful.

On the other hand, some searchers monitor the mempool for MEV-related transactions and attempt to preempt them, or might deliberately execute a transaction that generates an MEV opportunity and later execute another transaction to capture that MEV. For example, a searcher might detect a potential victim transaction aimed at exploiting an arbitrage opportunity, copy it, and then resubmit it with a higher gas fee to capture the MEV before the original transaction is executed. Such transactions are referred to as "value-diverting" MEV transactions because they divert value from victims to the searcher \cite{mohan2024blockchains}. Such transactions can be considered harmful, as they negatively impact other users. Value-diverting transactions have a detrimental impact on the DeFi system, especially when multiple searchers engage in open-bid first-price\footnote{In a first-price auction, the winning searcher pays their bid amount.} on-chain auctions to capture MEV. These competitive auctions, known as Priority Gas Auctions (PGAs) \cite{daian2020flash}, force searchers to pay significantly higher gas fees to prioritize their transactions. This bidding war increases the average gas price across the network, discouraging honest users from utilizing the platform. Additionally, searchers often submit multiple redundant transactions with progressively higher gas fees to maintain their position in the auction, leading to network congestion and increased latency \cite{mohan2024blockchains}. Daian et al. \cite{daian2020flash} illustrates an example of a PGA between two searchers, where the winning searcher issued 42 transactions consuming 113,265 gas, while the losing searcher issued 43 transactions consuming 33,547 gas. Further details on which MEV transactions are value-creating and which are value-diverting are explained in Section \ref{MEV_transactions}.

\section{Related Survey}
Several studies have examined different types of MEV transactions and their mitigation strategies within the DeFi ecosystem \cite{li2022survey,alipanahloo2024maximal,mohan2024blockchains,poux2022maximal,rasheed2024mevecosystem,gramlich2024maximal}. Li et al. \cite{li2022survey} analyzed vulnerabilities, attacks, and security optimizations in DeFi across different blockchain layers, noting that the consensus layer is susceptible to front-running and sandwich transactions that contribute to MEV. However, the survey lacks a thorough analysis of MEV transactions and does not address detection approaches or mitigation strategies.

Alipanahloo et al. \cite{alipanahloo2024maximal} categorized MEV mitigation strategies based on their potential to reduce or democratize MEV. They highlighted that fair transaction ordering policies, encrypted mempool and private transactions, and smart contract-based strategies are effective in reducing MEV, while the Proposer-Builder Separation (PBS) approach serves to democratize MEV. The study also assessed the impact of these strategies on network performance and implementation challenges. The authors find that mempool privacy through delay encryption and PBS strategies show promise in countering unfair MEV opportunities. Mohan and Khezr \cite{mohan2024blockchains} presented a temporal evolution of Ethereum protocols driven by the underlying need for MEV mitigation. However, their focus is limited to mitigation strategies based on private transactions.  However, these studies \cite{mohan2024blockchains,alipanahloo2024maximal} do not consider MEV transactions, detection approaches, and simulation methods.

Poux et al. \cite{poux2022maximal} provided a categorization of MEV transactions based on fairness and analyzed their impact on the network. They highlighted that liquidation and arbitrage transactions have a positive impact, whereas front-running, sandwich, and time-bandit transactions negatively affect the system. Furthermore, their study briefly discussed the private transactions-based MEV mitigation strategies. Rasheed et al. \cite{rasheed2024mevecosystem} analyzed front-running, back-running, and sandwich MEV transactions along with the evolution of private transactions-based MEV mitigation strategies. Lastly, Gramlich et al. \cite{gramlich2024maximal} provided details on front-running, back-running, sandwich, arbitrage, and liquidation transactions, along with various countermeasures for mitigating MEV. While these studies \cite{poux2022maximal,rasheed2024mevecosystem, gramlich2024maximal} do not analyze the different types of MEV within each transaction type, nor do they explore detection approaches and simulation methods.

Table \ref{tab:related_survey} summarizes the focus of each study. As shown, most surveys concentrate on MEV mitigation strategies \cite{mohan2024blockchains,alipanahloo2024maximal,poux2022maximal,rasheed2024mevecosystem,gramlich2024maximal}, with \cite{li2022survey,poux2022maximal,rasheed2024mevecosystem,gramlich2024maximal} covering a few categories of MEV transactions. To our knowledge, no survey comprehensively addresses MEV from the perspectives of transaction types, detection approaches, mitigation strategies, and simulation methods. This survey aims to fill this gap.

\begin{table}[]
\centering
\caption{Summary of related surveys.}
\label{tab:related_survey}
\scalebox{0.9}{
\begin{tabular}{ccclclc}
\hline
\rowcolor[HTML]{EFEFEF} 
\multicolumn{1}{|c|}{\cellcolor[HTML]{EFEFEF}} & \multicolumn{6}{c|}{\cellcolor[HTML]{EFEFEF}MEV focus area} \\ \cline{2-7} 
\rowcolor[HTML]{EFEFEF} 
\multicolumn{1}{|c|}{\multirow{-2}{*}{\cellcolor[HTML]{EFEFEF}Work}} & \multicolumn{1}{c|}{\cellcolor[HTML]{EFEFEF}\shortstack{Transactions \\ taxonomy}} & \multicolumn{2}{c|}{\cellcolor[HTML]{EFEFEF}\shortstack{Detection\\ approaches}} & \multicolumn{2}{c|}{\cellcolor[HTML]{EFEFEF}\shortstack{Mitigation\\ strategies}} & \multicolumn{1}{c|}{\cellcolor[HTML]{EFEFEF}\shortstack{Simulation\\ methods}} \\ \hline

\multicolumn{1}{|c|}{\cite{li2022survey}} & \multicolumn{1}{c|}{$\circlelefthalfblack$} & \multicolumn{2}{c|}{$\mdwhtcircle$} & \multicolumn{2}{c|}{$\mdwhtcircle$} & \multicolumn{1}{c|}{$\mdwhtcircle$} \\ \hline

\multicolumn{1}{|c|}{\cite{mohan2024blockchains}} & \multicolumn{1}{c|}{$\mdwhtcircle$} & \multicolumn{2}{c|}{$\mdwhtcircle$} & \multicolumn{2}{c|}{$\circlelefthalfblack$} & \multicolumn{1}{c|}{$\mdwhtcircle$} \\ \hline

\multicolumn{1}{|c|}{\cite{alipanahloo2024maximal}} & \multicolumn{1}{c|}{$\mdwhtcircle$} & \multicolumn{2}{c|}{$\mdwhtcircle$} & \multicolumn{2}{c|}{$\mdblkcircle$} & \multicolumn{1}{c|}{$\mdwhtcircle$} \\ \hline

\multicolumn{1}{|c|}{\cite{poux2022maximal}} & \multicolumn{1}{c|}{$\circlelefthalfblack$} & \multicolumn{2}{c|}{$\mdwhtcircle$} & \multicolumn{2}{c|}{$\circlelefthalfblack$} & \multicolumn{1}{c|}{$\mdwhtcircle$} \\ \hline

\multicolumn{1}{|c|}{\cite{rasheed2024mevecosystem}} & \multicolumn{1}{c|}{$\circlelefthalfblack$} & \multicolumn{2}{c|}{$\mdwhtcircle$} & \multicolumn{2}{c|}{$\circlelefthalfblack$} & \multicolumn{1}{c|}{$\mdwhtcircle$} \\ \hline

\multicolumn{1}{|c|}{\cite{gramlich2024maximal}} & \multicolumn{1}{c|}{$\circlelefthalfblack$} & \multicolumn{2}{c|}{$\mdwhtcircle$} & \multicolumn{2}{c|}{$\mdblkcircle$} & \multicolumn{1}{c|}{$\mdwhtcircle$} \\ \hline

\multicolumn{1}{|c|}{This survey} & \multicolumn{1}{c|}{$\mdblkcircle$} & \multicolumn{2}{c|}{$\mdblkcircle$} & \multicolumn{2}{c|}{$\mdblkcircle$} & \multicolumn{1}{c|}{$\mdblkcircle$} \\ \hline

\multicolumn{5}{l}{\footnotesize$\mdblkcircle$: considered; $\mdwhtcircle$: not considered; $\circlelefthalfblack$: partially considered}
\end{tabular}
}
\end{table}

\section{Methodology}
This section details the methodology employed to conduct the survey. In particular, we describe the literature retrieval strategy, the inclusion and exclusion criteria, and the article selection approach.

\subsection{Literature Retrieval}
This step involves selecting databases, formulating search strings, and retrieving articles. To conduct a comprehensive survey, we queried multiple databases, including the ACM Digital Library, IEEE Xplore, ScienceDirect, Scopus, Springer, Taylor \& Francis, and Web of Science. We developed search strings focusing on "Maximal/Miner Extractable Value" and "Ethereum/Smart Contract" to extract relevant articles. The initial search was limited to publications up to and including May 20, 2024. Following the initial database search, we performed an additional manual search on Google Scholar to identify articles that were not extracted via database search. This approach ensures an exhaustive exploration of the literature.

\subsection{Inclusion and Exclusion Criteria}
This step involves evaluating the records retrieved from the previous step to determine their suitability for inclusion in the survey based on specific inclusion and exclusion criteria. The inclusion criteria are as follows: \circled{1} The study is published in the English language, \circled{2} The research primarily focuses on MEV in the Ethereum network, and \circled{3} The study proposes an MEV detection approach, mitigation strategy, or simulation/extraction method. The exclusion criteria are: \circled{1} The study is a review or survey, \circled{2} The study is a book, \circled{3} The research focuses solely on the Ethereum platform without addressing MEV, and \circled{4} The study does not cover MEV detection, mitigation, or simulation.

\subsection{Article Selection}
This step involves selecting articles for inclusion in the survey. We followed the Preferred Reporting Items for Systematic Reviews and Meta-Analyses (PRISMA) 2020 guidelines \cite{page2021prisma} for article selection. A total of 215 records were retrieved through the database searches, along with 21 records from the manual search. Initially, 77 duplicate records were removed, leaving 159 records for the screening stage. These records were screened based on their titles and abstracts, resulting in the exclusion of 91 records, which reduced the total number of articles to 68. A full-text analysis of these 68 articles was then conducted to identify the most relevant ones for inclusion in the survey. Applying the exclusion criteria, 21 articles were excluded, leading to the inclusion of 47 articles in this survey

\section{Taxonomy of MEV Transactions} \label{MEV_transactions}

\begin{figure*}
    \centering
    \includegraphics[width=1\linewidth]{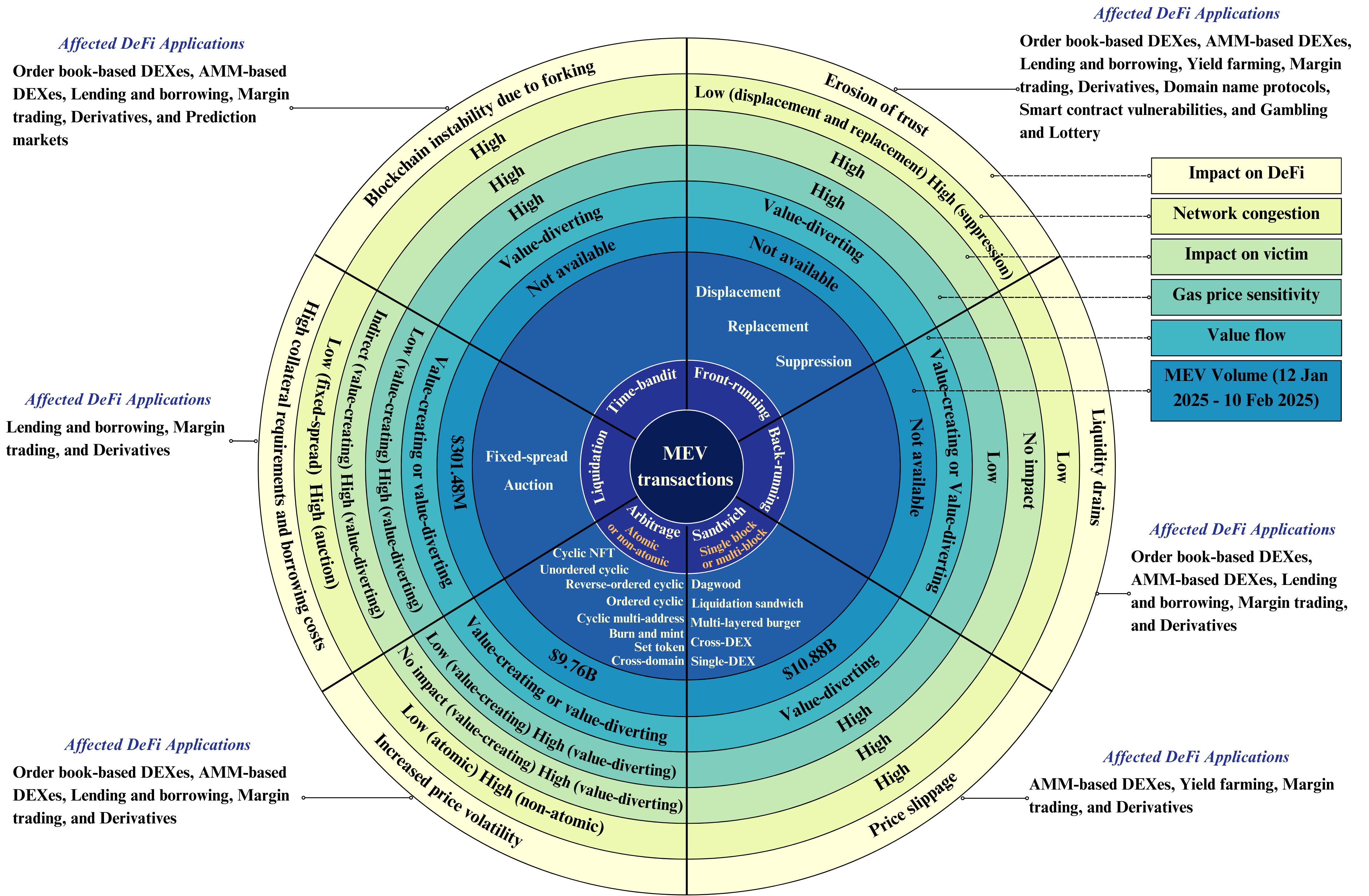}
    \caption{Taxonomy and comparison of MEV transactions.}
    \label{fig:taxonomy}
\end{figure*}

MEV transactions can be classified into several types based on the transaction details (such as token transfer and add/remove liquidity) and/or their order within a block relative to a victim transaction. Figure \ref{fig:taxonomy} presents our taxonomy of MEV transaction types and compares them across various metrics. In addition,  the figure depicts which DeFi applications are affected by each MEV transaction type. The analyses for gas price sensitivity and network congestion assume MEV transactions are executed by a single searcher without PGAs. The MEV volume statistics are obtained from EigenPhi\footnote{\url{https://eigenphi.io/}, accessed on 10 February 2025}. This section provides a detailed explanation of each transaction type, including real-world examples where applicable.

\subsection{Front-running}
In front-running, a searcher uses a victim transaction's information to submit their own transaction with a gas price higher than that of the victim. The front-run transaction is executed before the victim transaction, allowing the searcher to extract financial gains. Front-running transactions divert the value from the victim towards the searchers. These transactions could lead to increased gas prices, significant network congestion, unfair profits, and distrust towards Ethereum due to failed victim transactions \cite{momeni2022fairblock}. Front-running MEV transactions can be further categorized into displacement, replacement, and suppression.

\subsubsection{Displacement}
Displacement is the simplest form of front-running transactions, where a searcher displaces the victim transaction by executing their transaction first \cite{torres2021frontrunner}. Consider a scenario where a smart contract rewards the first user who submits a solution to a complex mathematical problem (e.g., finding twin prime chains or Cunningham chains). A victim transaction in the mempool includes the solution. A searcher spots this transaction, extracts the solution, and submits it with a higher gas fee (displacement transaction). The execution of the displacement transaction before the victim transaction diverts the reward toward the searcher. Figure \ref{fig:displacement} illustrates an example of a displacement front-running transaction.

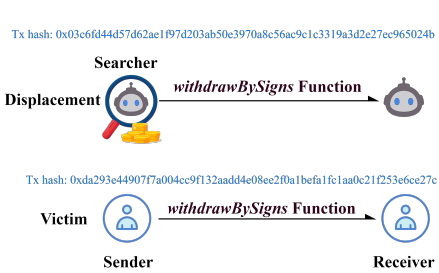
\begin{figure}
    \centering
    \resizebox{0.8\columnwidth}{!}{ 
        \begin{tikzpicture}[every node/.style={align=center, font=\footnotesize}]
        
        \node[anchor=west, text=blue, font=\ttfamily\normalsize] at (-8.5,5) {\href{https://etherscan.io/tx/0x03c6fd44d57d62ae1f97d203ab50e3970a8c56ac9c1c3319a3d2e27ec965024b}{Tx hash: 0x03c6fd44d57d62ae1f97d203ab50e3970a8c56ac9c1c3319a3d2e27ec965024b}};
        
        \draw[fill=purple!30, thick] (-5.8,3.2) rectangle (-3.2,4.2); 
        \node[font=\large] at (-4.5,3.7) {Searcher};
        \node[font=\large, anchor=east, scale=1.25] at (-6,3.7) {Displacement}; 
        
        \draw[fill=white!20, thick] (2.5,3.2) rectangle (5.5,4.2); 
        \node[font=\large] at (4,3.7) {Searcher};

        \draw[->, thick, >=latex, line width=0.5mm] (-3.2,3.7) -- (2.5,3.7); 
        \node[font=\large] at (-0.4,4.0) {\textit{withdrawBySigns} Function};

\node[anchor=west, text=blue, font=\ttfamily\normalsize] at (-8.5,2) {\href{https://etherscan.io/tx/0xda293e44907f7a004cc9f132aadd4e08ee2f0a1befa1fc1aa0c21f253e6ce27c}{Tx hash: 0xda293e44907f7a004cc9f132aadd4e08ee2f0a1befa1fc1aa0c21f253e6ce27c}};

\draw[fill=white!30, thick] (-5.8,0.2) rectangle (-3.2,1.2); 
\node[font=\large] at (-4.5,0.7) {Sender};
\node[font=\large, anchor=east, scale=1.25] at (-6,0.7) {Victim}; 

\draw[fill=white!20, thick] (2.5,0.2) rectangle (5.5,1.2); 
\node[font=\large] at (4,0.7) {Receiver};

\draw[->, thick, >=latex, line width=0.5mm] (-3.2,0.7) -- (2.5,0.7); 
\node[font=\large] at (-0.4,1.0) {\textit{withdrawBySigns} Function};

        \end{tikzpicture}
    }
    \caption{Displacement front-running transaction.}
    \label{fig:displacement}
\end{figure}

\subsubsection{Replacement}
Replacement front-running transactions replace the recipient address in a victim transaction to direct value to the searcher. On the Ethereum network, smart contracts often enable payments from a payer to a payee upon completing a specified performance obligation. Once the obligation is fulfilled, the payee initiates a transaction, providing the payment amount and proof of completion. However, a searcher can exploit this by monitoring the mempool for such transactions. The searcher then creates a replacement transaction, substituting the payee’s address with their own \cite{gans2023cryptography}. By offering a higher gas fee than the victim transaction, the replacement transaction is prioritized and executed first, diverting the payer’s funds to the searcher before the legitimate payee receives the payment.

\subsubsection{Suppression}
Suppression front-running transactions focus on congesting the network by repeatedly executing instructions to consume gas. This is to block a victim transaction from being included in a block \cite{torres2021frontrunner}. Suppression transactions could be executed either using an uncontrolled or controlled strategy. In an uncontrolled approach, the searcher lets the loop of instructions run until the block gas limit is reached and an out-of-gas exception is raised. In a controlled approach, the searcher monitors the gas usage and exits the loop just before the block gas limit is reached. Figure \ref{fig:suppression} depicts an example of a suppression front-running transaction.

\begin{figure}
    \centering
    \resizebox{0.8\columnwidth}{!}{ 
        \begin{tikzpicture}[every node/.style={align=center, font=\footnotesize}]

        \node[anchor=south, font=\large] at (-2, 5.6) {Block: 6207760}; 

        \draw[thick] (-9, 5.5) rectangle (7, -0.5); 

        \node[anchor=west, text=blue, font=\ttfamily\normalsize] at (-8.5,5) {\href{https://etherscan.io/tx/0x0bcda295549e5021b8b45f082637ce9ebc38937eb9e196b2614b82a11f4e1f8f}{Tx hash: 0x0bcda295549e5021b8b45f082637ce9ebc38937eb9e196b2614b82a11f4e1f8f}};
        
        \draw[fill=purple!30, thick] (-5.8,3.2) rectangle (-3.2,4.2); 
        \node[font=\large] at (-4.5,3.7) {Searcher};
        \node[font=\large, anchor=east, scale=1.25] at (-6,3.7) {Suppression}; 
        
        \draw[fill=white!20, thick] (2.5,3.2) rectangle (5.5,4.2); 
        \node[font=\large] at (4,3.7) {Contract};

        \draw[->, thick, >=latex, line width=0.5mm] (-3.2,3.7) -- (2.5,3.7); 
        \node[font=\large] at (-0.4,4.0) {\textit{repayBorrow} Function};

\node[anchor=west, text=blue, font=\ttfamily\normalsize] at (-8.5,2) {\href{https://etherscan.io/tx/0x001097c219905f2546b2f02450c87becbbf54ded904deccaf159a330c57da50d}{Tx hash: 0x001097c219905f2546b2f02450c87becbbf54ded904deccaf159a330c57da50d}};

\draw[fill=purple!30, thick] (-5.8,0.2) rectangle (-3.2,1.2); 
\node[font=\large] at (-4.5,0.7) {Searcher};
\node[font=\large, anchor=east, scale=1.25] at (-6,0.7) {Suppression}; 

\draw[fill=white!20, thick] (2.5,0.2) rectangle (5.5,1.2); 
\node[font=\large] at (4,0.7) {Contract};

\draw[->, thick, >=latex, line width=0.5mm] (-3.2,0.7) -- (2.5,0.7); 
\node[font=\large] at (-0.4,1.0) {\textit{repayBorrow} Function};

        \node[anchor=south, font=\large] at (-2, -1.4) {Block: 6207761}; 

     \draw[thick] (-9, -1.5) rectangle (7, -14); 

\node[anchor=west, text=blue, font=\ttfamily\normalsize] at (-8.5, -2.5) {\href{https://etherscan.io/tx/0x765b793d47d6b16c08b6d9405e111eced84f7987c5909e1edc7f1eec3f9dc565}{Tx hash: 0x765b793d47d6b16c08b6d9405e111eced84f7987c5909e1edc7f1eec3f9dc565}};

\draw[fill=purple!30, thick] (-5.8, -3.5) rectangle (-3.2, -4.5); 
\node[font=\large] at (-4.5, -4.0) {Searcher};
\node[font=\large, anchor=east, scale=1.25] at (-6, -4.0) {Suppression};

\draw[fill=white!20, thick] (2.5, -3.5) rectangle (5.5, -4.5); 
\node[font=\large] at (4, -4.0) {Contract};

\draw[->, thick, >=latex, line width=0.5mm] (-3.2, -4.0) -- (2.5, -4.0);
\node[font=\large] at (-0.4, -3.7) {\textit{repayBorrow} Function};

\node[anchor=west, text=blue, font=\ttfamily\normalsize] at (-8.5, -5.5) {\href{https://etherscan.io/tx/0xf75a34856feed1e692a4a1d68886104ac5a7b2e4116919e7a7e7d16eaca2fee6}{Tx hash: 0xf75a34856feed1e692a4a1d68886104ac5a7b2e4116919e7a7e7d16eaca2fee6}};

\draw[fill=purple!30, thick] (-5.8, -6.5) rectangle (-3.2, -7.5); 
\node[font=\large] at (-4.5, -7.0) {Searcher};
\node[font=\large, anchor=east, scale=1.25] at (-6, -7.0) {Suppression};

\draw[fill=white!20, thick] (2.5, -6.5) rectangle (5.5, -7.5); 
\node[font=\large] at (4, -7.0) {Contract};

\draw[->, thick, >=latex, line width=0.5mm] (-3.2, -7.0) -- (2.5, -7.0);
\node[font=\large] at (-0.4, -6.7) {\textit{repayBorrow} Function};

\node[anchor=west, text=blue, font=\ttfamily\normalsize] at (-8.5, -8.5) {\href{https://etherscan.io/tx/0xd583e7e8b43b3dd89cf4ffbcc17846c0691531f8d1ece92701c0e074bc283039}{Tx hash: 0xd583e7e8b43b3dd89cf4ffbcc17846c0691531f8d1ece92701c0e074bc283039}};

\draw[fill=purple!30, thick] (-5.8, -9.5) rectangle (-3.2, -10.5); 
\node[font=\large] at (-4.5, -10.0) {Searcher};
\node[font=\large, anchor=east, scale=1.25] at (-6, -10.0) {Suppression};

\draw[fill=white!20, thick] (2.5, -9.5) rectangle (5.5, -10.5); 
\node[font=\large] at (4, -10.0) {Contract};

\draw[->, thick, >=latex, line width=0.5mm] (-3.2, -10.0) -- (2.5, -10.0);
\node[font=\large] at (-0.4, -9.7) {\textit{repayBorrow} Function};

\node[anchor=west, text=blue, font=\ttfamily\normalsize] at (-8.5, -11.5) {\href{https://etherscan.io/tx/0x029f3f590bb00fa21e45879a308f70a2c4199d5eb4ac1007b9d89c772d5ab05b}{Tx hash: 0x029f3f590bb00fa21e45879a308f70a2c4199d5eb4ac1007b9d89c772d5ab05b}};

\draw[fill=purple!30, thick] (-5.8, -12.5) rectangle (-3.2, -13.5); 
\node[font=\large] at (-4.5, -13.0) {Searcher};
\node[font=\large, anchor=east, scale=1.25] at (-6, -13.0) {Suppression};

\draw[fill=white!20, thick] (2.5, -12.5) rectangle (5.5, -13.5); 
\node[font=\large] at (4, -13.0) {Contract};

\draw[->, thick, >=latex, line width=0.5mm] (-3.2, -13.0) -- (2.5, -13.0);
\node[font=\large] at (-0.4, -12.7) {\textit{repayBorrow} Function};

        \end{tikzpicture}
    }
    \caption{Suppression front-running transaction.}
    \label{fig:suppression}
\end{figure}

\subsection{Back-running}
In back-running, a searcher monitors the mempool for a victim transaction that causes a price discrepancy, which can be arbitraged. The searcher then submits a back-running transaction after the victim transaction, at a lower gas price, to capture the price difference. Back-running can be value-creating by reducing price discrepancies and contributing to market stability. However, it reduces profitable opportunities (value-diverting) for subsequent users due to liquidity drain, particularly for those who execute opposite-direction trades. Figure \ref{fig:backrunning} presents an example of a back-running transaction.

\begin{figure}
    \centering
    \resizebox{0.8\columnwidth}{!}{ 
        \begin{tikzpicture}[every node/.style={align=center, font=\footnotesize}]
        
        \node[anchor=west, text=blue, font=\ttfamily\normalsize] at (-8.5,5) {\href{https://etherscan.io/tx/0x304c15b1d93ce370cdb84ff1412b14a6cc932658c24dfe125b4344af7e226a39}{Tx hash: 0x304c15b1d93ce370cdb84ff1412b14a6cc932658c24dfe125b4344af7e226a39}};
        
\draw[fill=white!30, thick] (-5.8, 3.2) rectangle (-3.3, 4.2); 
\node[font=\large] at (-4.5, 3.7) {Sender}; 
\node[font=\large, anchor=east, scale=1.25] at (-6, 3.7) {Victim}; 

\draw[fill=white!20, thick] (2.5,3.2) rectangle (5.5,4.2); 
\node[font=\large] at (3.9,3.7) {Uniswap V2};

\draw[->, thick, >=latex, line width=0.5mm] (-3.3,3.8) -- (2.5,3.8); 
\draw[<-, thick, >=latex, line width=0.5mm] (-3.3,3.6) -- (2.5,3.6); 
\node[font=\large] at (-0.5,3.7) {130751.72188 NEIRO\\[0.2cm]5.75897 ETH}; 

\node[anchor=west, text=blue, font=\ttfamily\normalsize] at (-8.5,2) {\href{https://etherscan.io/tx/0xd41e0e2585413bc0d5c64a18f110d533bda84fa00fb76b71252e22ea00e5edd1}{Tx hash: 0xd41e0e2585413bc0d5c64a18f110d533bda84fa00fb76b71252e22ea00e5edd1}};

\draw[fill=purple!30, thick] (-5.8,0.2) rectangle (-3.2,1.2); 
\node[font=\large] at (-4.5,0.7) {Searcher};
\node[font=\large, anchor=east, scale=1.25] at (-6,0.7) {Back-run}; 

\draw[fill=white!20, thick] (2.5,0.2) rectangle (5.5,1.2); 
\node[font=\large] at (4,0.7) {Uniswap V2};

\draw[<-, thick, >=latex, line width=0.5mm] (-3.2,0.7) -- (2.5,0.7); 
\draw[->, thick, >=latex, line width=0.5mm] (-3.2,0.9) -- (2.5,0.9); 
\node[font=\large] at (-0.5,0.8) {10.10620 ETH\\[0.2cm]225626.30336 NEIRO}; 

        \end{tikzpicture}
    }
    \caption{Back-running transaction.}
    \label{fig:backrunning}
\end{figure}

\subsection{Sandwich}
A sandwich transaction typically involves both front-running and back-running transactions. In this, a searcher monitors the mempool for a victim transaction involving a large token swap and then initiates a front-run transaction with the same swap but at a higher gas price. When the searcher's transaction is executed, the token exchange price temporarily rises, causing the victim to swap tokens at a higher price than expected. After the victim's transaction, the searcher performs a back-run transaction to reverse the earlier front-run swap. The searcher profits from the difference between the token exchange price the victim expected to pay and the price actually paid. Sandwich transactions are considered as value-diverting MEV transactions \cite{mohan2024blockchains} and lead to financial losses for users conducting token trades due to price slippage. Additionally, each sandwich attempt introduces two extra transactions (front-run and back-run) into the network, leading to increased block space utilization \cite{poux2022maximal} and bloating of mempool, which in turn reduces blockchain efficiency. Sandwich transactions could occur either in a single block or could span across multiple blocks. Below, we describe different types of sandwich transactions used to extract MEV, providing examples of single-block scenarios.

\subsubsection{Single DEX}
This is the simplest form of a sandwich transaction, where the front-run, victim, and back-run transactions all occur on a single DEX \cite{park2024unraveling}. A searcher initiates a swap transaction identical to the victim's and front-runs it by paying a higher gas fee than the victim's transaction. Following this, the searcher executes a reverse swap and back-runs the victim transaction by paying a lower gas fee. Figure \ref{fig:sandwich_single_dex} illustrates an example of a single DEX sandwich transaction.

\begin{figure}
    \centering
    \resizebox{0.8\columnwidth}{!}{ 
        \begin{tikzpicture}[every node/.style={align=center, font=\footnotesize}]
        
\node[anchor=west, text=blue, font=\ttfamily\normalsize] at (-8.5,5) {\href{https://etherscan.io/tx/0xfe832ab9351a516c7a56501b0555b65d62d10c76d04e8544a7f20915406ecacc}{Tx hash: 0xfe832ab9351a516c7a56501b0555b65d62d10c76d04e8544a7f20915406ecacc}};

\draw[fill=purple!30, thick] (-5.8,3.2) rectangle (-3.2,4.2); 
\node[font=\large] at (-4.5,3.7) {Searcher};
\node[font=\large, anchor=east, scale=1.25] at (-6,3.7) {Front-run}; 

\draw[fill=white!20, thick] (2.5,3.2) rectangle (5.5,4.2); 
\node[font=\large] at (4,3.7) {Uniswap V2};

\draw[<-, thick, >=latex, line width=0.5mm] (-3.2,3.7) -- (2.5,3.7); 
\draw[->, thick, >=latex, line width=0.5mm] (-3.2,3.9) -- (2.5,3.9); 
\node[font=\large] at (-0.5,3.8) {4.43166 WETH\\[0.2cm]3390220985.74 ONIGI}; 

\node[anchor=west, text=blue, font=\ttfamily\normalsize] at (-8.5,2) {\href{https://etherscan.io/tx/0x89d7b6a885e1dcb0a579e6ae69d74f9d7092984a5c8aa3f4981905298882cba3}{Tx hash: 0x89d7b6a885e1dcb0a579e6ae69d74f9d7092984a5c8aa3f4981905298882cba3}};

\draw[fill=white!30, thick] (-5.8, 0.2) rectangle (-3.3, 1.2); 
\node[font=\large] at (-4.5, 0.7) {Sender}; 
\node[font=\large, anchor=east, scale=1.25] at (-6, 0.7) {Victim}; 

\draw[fill=white!20, thick] (2.5, 0.2) rectangle (5.5, 1.2); 
\node[font=\large] at (3.9, 0.7) {Uniswap V2}; 

\draw[->, thick, >=latex, line width=0.5mm] (-3.3, 0.8) -- (2.5, 0.8); 
\draw[<-, thick, >=latex, line width=0.5mm] (-3.3, 0.6) -- (2.5, 0.6); 
\node[font=\large] at (-0.5, 0.7) {0.327 WETH\\[0.2cm]223226622.09 ONIGI}; 

\node[anchor=west, text=blue, font=\ttfamily\normalsize] at (-8.5,-1) {\href{https://etherscan.io/tx/0x130c1887662105cf87fee420ce1c7b931c55cb97fead899256ccae0e6133efd3}{Tx hash: 0x130c1887662105cf87fee420ce1c7b931c55cb97fead899256ccae0e6133efd3}};

\draw[fill=purple!30, thick] (-5.8,-2.8) rectangle (-3.2,-1.8); 
\node[font=\large] at (-4.5,-2.3) {Searcher};
\node[font=\large, anchor=east, scale=1.25] at (-6,-2.3) {Back-run}; 

\draw[fill=white!20, thick] (2.5,-2.8) rectangle (5.5,-1.8); 
\node[font=\large] at (4,-2.3) {Uniswap V2};

\draw[<-, thick, >=latex, line width=0.5mm] (-3.2,-2.3) -- (2.5,-2.3); 
\draw[->, thick, >=latex, line width=0.5mm] (-3.2,-2.1) -- (2.5,-2.1); 
\node[font=\large] at (-0.5,-2.2) {3390220150.37 ONIGI\\[0.2cm]4.47029 WETH}; 
        \end{tikzpicture}
    }
    \caption{Single DEX sandwich transaction.}
    \label{fig:sandwich_single_dex}
\end{figure}
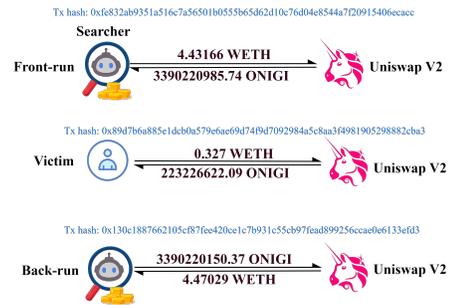

\subsubsection{Cross-DEX}
In cross-DEX sandwich transactions, a searcher usually lacks the necessary tokens to perform the same swap transaction as the victim \cite{park2024unraveling}. To acquire the required tokens, the searcher first completes a transaction on a DEX different from the one where the victim is swapping tokens. Once the tokens are obtained, the searcher executes a front-run transaction on the same DEX as the victim. After the victim transaction is processed, the searcher performs a back-run transaction to reverse the front-run trade and secure the sandwich profit. Additionally, the searcher may also seek arbitrage profit by exploiting the price difference between the two exchanges. To do this, the searcher may execute another transaction on the original DEX to reverse the initial trade used to acquire the tokens, thereby securing the arbitrage profit. Figure \ref{fig:sandwich_cross_dex} illustrates an example of a cross-DEX sandwich transaction.

\begin{figure}
    \centering
    \resizebox{0.8\columnwidth}{!}{ 
        \begin{tikzpicture}[every node/.style={align=center, font=\footnotesize}]
        
        \node[anchor=west, text=blue, font=\ttfamily\normalsize] at (-8.5, 6.4) {\href{https://etherscan.io/tx/0x3ba534712841c5cf8b03df6c72fc0efdca24e010ebb845c5cfba90a0cbd529f7}{Tx hash: 0x3ba534712841c5cf8b03df6c72fc0efdca24e010ebb845c5cfba90a0cbd529f7}};
        
        \draw[fill=purple!30, thick] (-5.8,2.1) rectangle (-3.2,3.1); 
        \node[font=\large] at (-4.5,2.6) {Searcher};
        \node[font=\large, anchor=east, scale=1.25] at (-6,2.5) {Front-run}; 

        \draw[fill=white!20, thick] (2.5,4.7) rectangle (5.5,5.7); 
        \node[font=\large] at (4,5.2) {Uniswap V3};
        
\draw[fill=white!20, thick] (2.5,2.1) rectangle (5.5,3.1); 
\node[font=\large] at (4,2.6) {Uniswap V2}; 
        \draw[->, thick, >=latex, line width=0.5mm] (-3.2,2.9) -- (2.5,5.4);
        \draw[<-, thick, >=latex, line width=0.5mm] (-3.2,2.7) -- (2.5,5.2);
\node[font=\large, rotate=22] at (-0.2, 4.1) {0.753554 WETH\\[0.2cm]1936.94 USDT}; 

\draw[<-, thick, >=latex, line width=0.5mm] (2.5,2.6) -- (-3.2,2.6); 
\draw[->, thick, >=latex, line width=0.5mm] (2.5,2.4) -- (-3.2,2.4); 
\node[font=\large] at (-0.2,2.5) {1936.94 USDT\\[0.2cm]19345.56 Vow}; 

\node[anchor=west, text=blue, font=\ttfamily\normalsize] at (-8.5,1) {\href{https://etherscan.io/tx/0xc87e5d682e5f8b13b9b5f3f1d0017540a7d23417c1e09e6fda7ccdd2ead166bd}{Tx hash: 0xc87e5d682e5f8b13b9b5f3f1d0017540a7d23417c1e09e6fda7ccdd2ead166bd}};

\draw[fill=white!30, thick] (-5.8,-0.8) rectangle (-3.2,0.2); 
\node[font=\large] at (-4.5,-0.3) {Sender};
\node[font=\large, anchor=east, scale=1.25] at (-6,-0.3) {Victim}; 

\draw[fill=white!20, thick] (2.5,-0.8) rectangle (5.5,0.2); 
\node[font=\large] at (4,-0.3) {Uniswap V2};

\draw[->, thick, >=latex, line width=0.5mm] (-3.2,-0.3) -- (2.5,-0.3); 
\draw[<-, thick, >=latex, line width=0.5mm] (-3.2,-0.5) -- (2.5,-0.5); 
\node[font=\large] at (-0.2,-0.4) {3421.41 USDT\\[0.2cm]33700.34 Vow}; 

\node[anchor=west, text=blue, font=\ttfamily\normalsize] at (-8.5, -2.6) {\href{https://etherscan.io/tx/0x39fdb17033b2b9572aa43e075337bf7258c51a1e2c2fcf1ccf854cf58f416e57}{Tx hash: 0x39fdb17033b2b9572aa43e075337bf7258c51a1e2c2fcf1ccf854cf58f416e57}};
        
\draw[fill=purple!30, thick] (-5.8,-4.3) rectangle (-3.2,-3.3); 
\node[font=\large] at (-4.5,-3.8) {Searcher};
\node[font=\large, anchor=east, scale=1.25] at (-6,-3.8) {Back-run}; 

\draw[fill=white!20, thick] (2.5,-4.3) rectangle (5.5,-3.3); 
\node[font=\large] at (4,-3.8) {Uniswap V2};
        
\draw[fill=white!20, thick] (2.5,-7.3) rectangle (5.5,-6.3); 
\node[font=\large] at (4,-6.8) {Uniswap V3}; 

\draw[->, thick, >=latex, line width=0.5mm] (-3.2,-3.6) -- (2.5,-3.6);
\draw[<-, thick, >=latex, line width=0.5mm] (-3.2,-3.8) -- (2.5,-3.8);
\node[font=\large] at (-0.2,-3.7) {19345.56     Vow\\[0.2cm]1959.71 USDT}; 

\draw[<-, thick, >=latex, line width=0.5mm] (2.5,-6.7) -- (-3.2,-4.0); 
\draw[->, thick, >=latex, line width=0.5mm] (2.5,-6.9) -- (-3.2,-4.2); 
\node[font=\large, rotate=-25] at (-0.2,-5.5) {1959.71 USDT\\[0.2cm]0.7622577 WETH}; 

        \end{tikzpicture}
    }
    \caption{Cross-DEX sandwich transaction.}
    \label{fig:sandwich_cross_dex}
\end{figure}
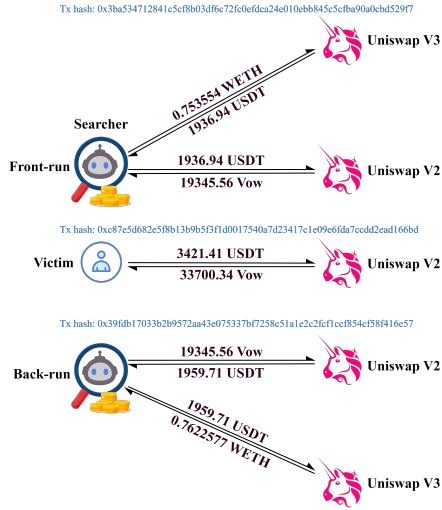

\subsubsection{Multi-layered burger}
In multi-layer burger sandwich transactions, a searcher exploits the information of multiple victim transactions to maximize their profit \cite{li2023demystifying}. The searcher looks for multiple large transactions with the same swap and front-runs those transactions. After the execution of victim transactions, the searcher performs a back-run transaction to trade the previously swapped tokens. The profit in this type of transaction potentially increases with more victim transactions. This is because each token swap transaction by a victim will impact the token exchange price in the searcher's favor. Figure \ref{fig:sandwich_multi_victim} presents an example of a multi-victim burger sandwich transaction. This type of transaction could be single DEX or cross-DEX.

\begin{figure}
    \centering
    \resizebox{0.8\columnwidth}{!}{ 
        \begin{tikzpicture}[every node/.style={align=center, font=\footnotesize}]
        
\node[anchor=west, text=blue, font=\ttfamily\normalsize] at (-8.5,5) {\href{https://etherscan.io/tx/0xc1b22d28827d31746cc923fdb90b54de7efdec5fcb3049531d591bffd2ee9060}{Tx hash: 0xc1b22d28827d31746cc923fdb90b54de7efdec5fcb3049531d591bffd2ee9060}};

\draw[fill=purple!30, thick] (-5.8,3.2) rectangle (-3.2,4.2); 
\node[font=\large] at (-4.5,3.7) {Searcher};
\node[font=\large, anchor=east, scale=1.25] at (-6,3.7) {Front-run}; 

\draw[fill=white!20, thick] (2.5,3.2) rectangle (5.5,4.2); 
\node[font=\large] at (4,3.7) {Uniswap V2};

\draw[<-, thick, >=latex, line width=0.5mm] (-3.2,3.7) -- (2.5,3.7); 
\draw[->, thick, >=latex, line width=0.5mm] (-3.2,3.9) -- (2.5,3.9); 
\node[font=\large] at (-0.5,3.8) {0.3332138 WETH\\[0.2cm]235317.26 TAONU}; 

\node[anchor=west, text=blue, font=\ttfamily\normalsize] at (-8.5,2) {\href{https://etherscan.io/tx/0x2e529c4cc48ca3425b4481910114294b3e5482a0c82d6881540009d7d1b6290a}{Tx hash: 0x2e529c4cc48ca3425b4481910114294b3e5482a0c82d6881540009d7d1b6290a}};
\draw[fill=white!30, thick] (-5.8,0.2) rectangle (-3.2,1.2); 
\node[font=\large] at (-4.5,0.7) {Sender}; 
\node[font=\large, anchor=east, scale=1.25] at (-6,0.7) {Victim 1}; 

\draw[fill=white!20, thick] (2.5,0.2) rectangle (5.5,1.2); 
\node[font=\large] at (4,0.7) {Uniswap V2}; 

\draw[->, thick, >=latex, line width=0.5mm] (-3.2,0.7) -- (2.5,0.7); 
\draw[<-, thick, >=latex, line width=0.5mm] (-3.2,0.5) -- (2.5,0.5); 
\node[font=\large] at (-0.2,0.6) {0.3721647 WETH\\[0.2cm]260071.3 TAONU}; 

\node[anchor=west, text=blue, font=\ttfamily\normalsize] at (-8.5,-1) {\href{https://etherscan.io/tx/0xcd72252e3f07deae7571842ac573d912e96ced44257527d31390a33f1295e4e3}{Tx hash: 0xcd72252e3f07deae7571842ac573d912e96ced44257527d31390a33f1295e4e3}};

\draw[fill=white!30, thick] (-5.8,-2.8) rectangle (-3.2,-1.8); 
\node[font=\large] at (-4.5,-2.3) {Sender}; 
\node[font=\large, anchor=east, scale=1.25] at (-6,-2.3) {Victim 2}; 

\draw[fill=white!20, thick] (2.5,-2.8) rectangle (5.5,-1.8); 
\node[font=\large] at (4,-2.3) {Uniswap V2}; 

\draw[->, thick, >=latex, line width=0.5mm] (-3.2,-2.3) -- (2.5,-2.3); 
\draw[<-, thick, >=latex, line width=0.5mm] (-3.2,-2.5) -- (2.5,-2.5); 
\node[font=\large] at (-0.2,-2.4) {0.186165 WETH\\[0.2cm]129017.03 TAONU}; 

\node[anchor=west, text=blue, font=\ttfamily\normalsize] at (-8.5,-4) {\href{https://etherscan.io/tx/0x9fc5cb6b3ffe47e2687485b27cff8097315de1500603b67aa96ccfeb34e885af}{Tx hash: 0x9fc5cb6b3ffe47e2687485b27cff8097315de1500603b67aa96ccfeb34e885af}};

\draw[fill=purple!30, thick] (-5.8,-5.8) rectangle (-3.2,-4.8); 
\node[font=\large] at (-4.5,-5.3) {Searcher};
\node[font=\large, anchor=east, scale=1.25] at (-6,-5.3) {Back-run}; 

\draw[fill=white!20, thick] (2.5,-5.8) rectangle (5.5,-4.8); 
\node[font=\large] at (4,-5.3) {Uniswap V2};

\draw[<-, thick, >=latex, line width=0.5mm] (-3.2,-5.3) -- (2.5,-5.3); 
\draw[->, thick, >=latex, line width=0.5mm] (-3.2,-5.1) -- (2.5,-5.1); 
\node[font=\large] at (-0.5,-5.2) {235317.19 TAONU \\[0.2cm]0.3367603 WETH}; 
        \end{tikzpicture}
    }
    \caption{Multi-layered burger sandwich transaction.}
    \label{fig:sandwich_multi_victim}
\end{figure}
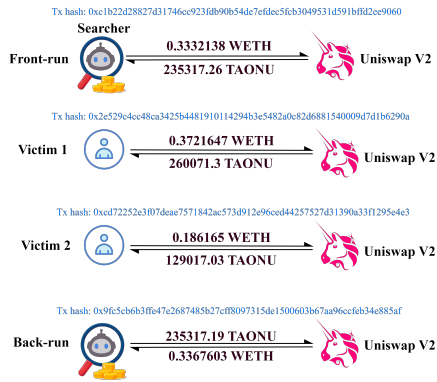

\subsubsection{Dagwood}
In Dagwood sandwich transactions, a searcher executes multiple front-running transactions followed by a single back-running transaction to extract MEV \cite{bartoletti2022maximizing}. This type of transaction is also known as a conjoined sandwich \cite{chi2024remeasuring}. A Dagwood sandwich can involve either a single victim transaction placed between the two front-running or front-running and back-running transactions \cite{bartoletti2022maximizing}, or it can involve multiple victim transactions sandwiched between these transactions \cite{wang2023n}.

\subsubsection{Liquidation sandwich}
In the sandwich transaction types explained above, a searcher performs token transfer front-run and back-run transactions to profit from price slippage. However, in a liquidation sandwich transaction, a searcher performs liquidation front-run and back-run transactions to profit from the exchange fee acquired from the victim's swap transaction \cite{li2023demystifying, xiong2023demystifying}. In this transaction, a searcher acts as a liquidity provider on the same DEX where the victim swaps tokens. This is achieved by performing a front-run transaction to supply the tokens to be swapped and a back-run transaction to withdraw the swapped tokens. Figure \ref{fig:liquidation_sandwich} represents an example of a liquidation sandwich transaction.

\begin{figure}
    \centering
    \resizebox{0.8\columnwidth}{!}{ 
        \begin{tikzpicture}[every node/.style={align=center, font=\footnotesize}]
        
\node[anchor=west, text=blue, font=\ttfamily\normalsize] at (-8.5,5) {\href{https://etherscan.io/tx/0xe824b92269f43bc4afd9e883ad38080b9388f2961fa0467f3334271f694bb371}{Tx hash: 0xe824b92269f43bc4afd9e883ad38080b9388f2961fa0467f3334271f694bb371}};

\draw[fill=purple!30, thick] (-5.8,3.2) rectangle (-3.2,4.2); 
\node[font=\large] at (-4.5,3.7) {Searcher};
\node[font=\large, anchor=east, scale=1.25] at (-6,3.7) {Front-run}; 

\draw[fill=white!20, thick] (2.5,3.2) rectangle (5.5,4.2); 
\node[font=\large] at (4,3.7) {Uniswap V3};


\draw[->, thick, >=latex, line width=0.5mm] (-3.2,3.7) -- (2.5,3.7); 
\node[font=\large] at (-0.5,3.8) {Add 11106782.94371 USDC \\ and 0 ETH\\[0.2cm]}; 

\node[anchor=west, text=blue, font=\ttfamily\normalsize] at (-8.5,2) {\href{https://etherscan.io/tx/0x34d66ff483426f41b245881a92ce450579742e29aa85f256b752ed2403e5c8d1}{Tx hash: 0x34d66ff483426f41b245881a92ce450579742e29aa85f256b752ed2403e5c8d1}};

\draw[fill=white!30, thick] (-5.8,0.2) rectangle (-3.2,1.2); 
\node[font=\large] at (-4.5,0.7) {Sender}; 
\node[font=\large, anchor=east, scale=1.25] at (-6,0.8) {Victim}; 

\draw[fill=white!20, thick] (2.5,0.2) rectangle (5.5,1.2); 
\node[font=\large] at (4,0.7) {Uniswap V3}; 

\draw[->, thick, >=latex, line width=0.5mm] (-3.2,0.8) -- (2.5,0.8); 
\draw[<-, thick, >=latex, line width=0.5mm] (-3.2,0.6) -- (2.5,0.6); 
\node[font=\large] at (-0.2,0.7) {1498.68164 ETH\\[0.2cm]6946663.72530 USDC}; 

\node[anchor=west, text=blue, font=\ttfamily\normalsize] at (-8.5,-1) {\href{https://etherscan.io/tx/0xaadd749a18b2b75f8eec5c12e8be4ccb2471ea3f96db8f7a76beab0090a03c84}{Tx hash: 0xaadd749a18b2b75f8eec5c12e8be4ccb2471ea3f96db8f7a76beab0090a03c84}};

\draw[fill=purple!30, thick] (-5.8,-2.8) rectangle (-3.2,-1.8); 
\node[font=\large] at (-4.5,-2.3) {Searcher};
\node[font=\large, anchor=east, scale=1.25] at (-6,-2.3) {Back-run}; 

\draw[fill=white!20, thick] (2.5,-2.8) rectangle (5.5,-1.8); 
\node[font=\large] at (4,-2.3) {Uniswap V3};

\draw[<-, thick, >=latex, line width=0.5mm] (-3.2,-2.3) -- (2.5,-2.3); 

\node[font=\large] at (-0.3,-2.2) {Remove 5542832.47128 USDC \\ and 1200.01436 ETH\\[0.2cm]}; 
        \end{tikzpicture}
    }
    \caption{Liquidation sandwich transaction.}
    \label{fig:liquidation_sandwich}
\end{figure}
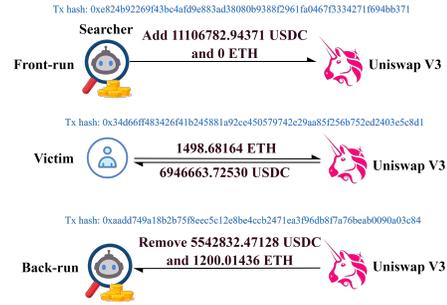

Although sandwich transactions are considered value-diverting MEV transactions and can result in financial losses for individual users, they might occasionally lead to improved overall network efficiency. Kulkarni et al. \cite{kulkarni2023routing} analyzed that a sandwich transaction adding liquidity routes could act as a decentralized controller, prompting users to avoid certain trading routes and thereby enhancing network efficiency. However, a more thorough investigation and study of real-world transactions are necessary to fully understand these effects. 

\subsection{Arbitrage}
Arbitrage transactions are the ones used by searchers to extract value by exploiting price slippage between asset pairs across multiple DEXes or by taking advantage of price differences between asset pairs within a single or multiple DEXes. Arbitrage transactions can be either value-creating or value-diverting. In a value-creating arbitrage scenario, a searcher identifies a price difference for an asset pair across different DEXes. They then execute an arbitrage transaction by swapping the assets on one DEX and performing a reverse swap on another. This type of arbitrage helps stabilize asset exchange prices across DEXes. Conversely, in a value-diverting scenario, a searcher monitors the mempool for pending high-profit arbitrage trades and front-runs these transactions by executing the arbitrage themselves. This can result in financial losses for the victim. Arbitrage transactions could be atomic (i.e., all swap events are performed using a single transaction) or non-atomic (i.e., multiple sequential transactions are performed to execute the swap events). Constant trading due to arbitrage transactions could lead to short-term price volatility, affecting market stability and predictability for other users. Below, we describe different types of arbitrage transactions used to gain MEV, providing examples of atomic scenarios.

\subsubsection{Ordered cyclic}
An ordered cyclic arbitrage transaction involves multiple swap events that form a cycle, where the input token for a swap is the same as the output token of the previous swap \cite{wang2022cyclic}. However, the values of the input and output tokens involved may differ. Furthermore, the output token for the last swap is the same as the input token for the first swap. This type of arbitrage can involve either a token pair (i.e., a transaction involving only two tokens) or token chains (i.e., multiple token pairs swapped within a single transaction). In an ordered cyclic transaction with a token pair, typically, two DEXes are used. The arbitrage exploits price slippage between DEXes to acquire MEV. This is often referred to as simple loop arbitrage \cite{park2024unraveling}. Conversely, an ordered cyclic transaction involving a token chain may include one or multiple DEXes. Figure \ref{fig:ordered_cyclic_token_pair} illustrates an example of an ordered cyclic arbitrage transaction using a token pair, while Figure \ref{fig:ordered_cyclic_token_chain} shows an example of an ordered cyclic arbitrage transaction involving a token chain. The numbers above each swap event in the figures represent the swap index within the transaction, indicating the sequence in which the swaps occur.

\begin{figure}
    \centering
    \resizebox{0.8\columnwidth}{!}{ 
        \begin{tikzpicture}[every node/.style={align=center, font=\footnotesize}]
        
\node[anchor=west, text=blue, font=\ttfamily\normalsize] at (-9.5, 4.9) {\href{https://etherscan.io/tx/0x2005a655b2158d1a9a0a310c85850a72068af95c52af96dd9d9df1c7caa0e64a}{Tx hash: 0x2005a655b2158d1a9a0a310c85850a72068af95c52af96dd9d9df1c7caa0e64a}};

\draw[fill=purple!30, thick] (-7.8, 2.1) rectangle (-5.2, 3.1); 
\node[font=\large] at (-6.5, 2.6) {Searcher};

\draw[fill=white!20, thick] (0.5, 3.2) rectangle (3.5, 4.2); 
\node[font=\large] at (2, 3.7) {Sushiswap};

\draw[fill=white!20, thick] (0.5, 1) rectangle (3.5, 2); 
\node[font=\large] at (2, 1.5) {Uniswap v3};

\draw[->, thick, >=latex, line width=0.5mm] (-5.2, 2.9) -- (0.5, 3.8);
\draw[<-, thick, >=latex, line width=0.5mm] (-5.2, 2.7) -- (0.5, 3.6);
\node[font=\large, rotate=10] at (-1.8, 3.35) {0.55533 ETH\\[0.2cm]557.70319 SYN};
\node[draw, circle, line width=0.1mm, fill=black, text=white, font=\small] at (-0.2, 4.2) {1}; 

\draw[<-, thick, >=latex, line width=0.5mm] (0.5, 1.6) -- (-5.2, 2.5);
\draw[->, thick, >=latex, line width=0.5mm] (0.5, 1.4) -- (-5.2, 2.3);
\node[font=\large, rotate=-10] at (-1.8, 1.85) {557.70319 SYN\\[0.2cm]0.56415 ETH};
\node[draw, circle, line width=0.1mm, fill=black, text=white, font=\small] at (-0.2, 2.45) {2};

        \end{tikzpicture}
    }
    \caption{Ordered cyclic arbitrage transaction using a token pair.}
    \label{fig:ordered_cyclic_token_pair}
\end{figure}
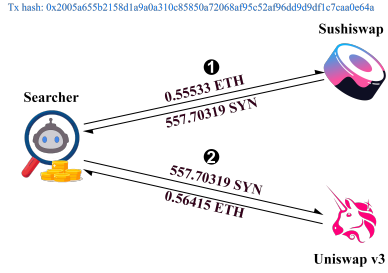

\begin{figure}
    \centering
    \resizebox{0.8\columnwidth}{!}{ 
        \begin{tikzpicture}[every node/.style={align=center, font=\footnotesize}]
        
\node[anchor=west, text=blue, font=\ttfamily\normalsize] at (-9.5, 8) {\href{https://etherscan.io/tx/0x4245a5e148ce0dcd65742eb354bfddfb9cd73df58f53a6aa6368e1f9ffa43044}{Tx hash: 0x4245a5e148ce0dcd65742eb354bfddfb9cd73df58f53a6aa6368e1f9ffa43044}};

\draw[fill=purple!30, thick] (-7.8, 2.0) rectangle (-5.2, 3.1); 
\node[font=\large] at (-6.5, 2.55) {Searcher};

\draw[fill=white!20, thick] (0.5, 6) rectangle (3.5, 7); 
\node[font=\large] at (2, 6.5) {Sushiswap};

\draw[fill=white!20, thick] (0.5, 3.8) rectangle (3.5, 4.8); 
\node[font=\large] at (2, 4.3) {Bancor};

\draw[fill=white!20, thick] (0.5, 1) rectangle (3.5, 2); 
\node[font=\large] at (2, 1.5) {Bancor};

\draw[fill=white!20, thick] (0.5, -1) rectangle (3.5, -2); 
\node[font=\large] at (2, -1.5) {Uniswap V2};

\draw[->, thick, >=latex, line width=0.5mm] (-5.2, 3.1) -- (0.5, 6.7); 
\draw[<-, thick, >=latex, line width=0.5mm] (-5.2, 2.9) -- (0.5, 6.5); 
\node[font=\large, rotate=32] at (-1.65, 5.25) {0.21 ETH\\[0.2cm]15008.20702 EDEN};
\node[draw, circle, line width=0.1mm, fill=black, text=white, font=\small] at (-0.8, 6.5) {1}; 

\draw[->, thick, >=latex, line width=0.5mm] (-5.2, 2.8) -- (0.5, 4.4);
\draw[<-, thick, >=latex, line width=0.5mm] (-5.2, 2.6) -- (0.5, 4.2);
\node[font=\large, rotate=17] at (-1.8, 3.65) {15008.20702 EDEN\\[0.2cm]879.92115 BNT};
\node[draw, circle, line width=0.1mm, fill=black, text=white, font=\small] at (-0.1, 4.9) {2};

\draw[<-, thick, >=latex, line width=0.5mm] (0.5, 1.6) -- (-5.2, 2.5);
\draw[->, thick, >=latex, line width=0.5mm] (0.5, 1.4) -- (-5.2, 2.35);
\node[font=\large, rotate=-10] at (-1.8, 1.85) {879.92115 BNT\\[0.2cm]147995.10490 BORING};

\node[draw, circle, line width=0.1mm, fill=black, text=white, font=\small] at (-0.1, 2.45) {3}; 

\draw[<-, thick, >=latex, line width=0.5mm] (0.5, -1.4) -- (-5.2, 2.3);
\draw[->, thick, >=latex, line width=0.5mm] (0.5, -1.6) -- (-5.2, 2.1);
\node[font=\large, rotate=-33] at (-1.7, -0.1) {147995.10490 BORING\\[0.2cm]0.23459 ETH};

\node[draw, circle, line width=0.1mm, fill=black, text=white, font=\small] at (-0.3, 0.1) {4}; 
        \end{tikzpicture}
    }
    \caption{Ordered cyclic arbitrage transaction using a token chain.}
    \label{fig:ordered_cyclic_token_chain}
\end{figure}
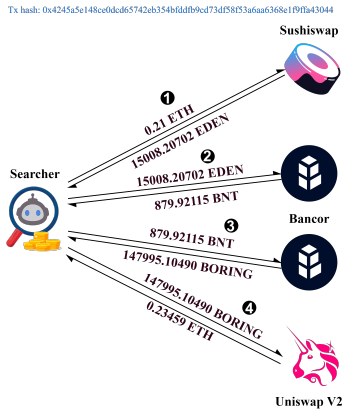

\subsubsection{Reverse-ordered cyclic}
A reverse-ordered cyclic arbitrage transaction is similar to an ordered cyclic arbitrage transaction, with the main difference being the sequence of swap events within the transaction. In this type of transaction, the output token of each swap is the same as the input token of the previous swap, although the values of the tokens involved may differ. Additionally, the output token of the first swap is the same as the input token of the last swap. Essentially, a reverse ordered cyclic arbitrage transaction is an ordered cyclic arbitrage transaction with the order of swap events reversed. Figure \ref{fig:reverse_ordered_cyclic} illustrates an example of a reverse-ordered cyclic arbitrage transaction.

\subsubsection{Unordered cyclic}
An unordered cyclic arbitrage transaction involves swap events where the input token for a swap is not the same as the output token of the previous swap \cite{chi2024remeasuring}. Although the token exchange graph forms a cycle, the swaps are not in a specific order. Despite this, the transaction still yields a profit. Figure \ref{fig:unordered_cyclic} illustrates an example of an unordered cyclic arbitrage transaction.

\subsubsection{Cyclic Non-Fungible Token (NFT)}
A cyclic NFT arbitrage transaction is similar to an ordered cyclic arbitrage transaction involving a token pair, with the key difference being that, instead of swapping between two ERC-20 tokens, a cyclic NFT arbitrage involves swapping a token with an NFT \cite{park2024unraveling}. Figure \ref{fig:cyclic_nft} illustrates an example of a cyclic NFT arbitrage transaction, where ETH is swapped for SUDOSQUID NFT with token ID 1587.

\subsubsection{Cyclic multi-address}
A cyclic multi-address arbitrage transaction involves more than one searcher address participating in different swap events \cite{park2024unraveling}. In this scenario, the extracted value for each address is aggregated to compute the total acquired MEV. Figure \ref{fig:cyclic_multi_address} illustrates an example of a cyclic multi-address arbitrage transaction.

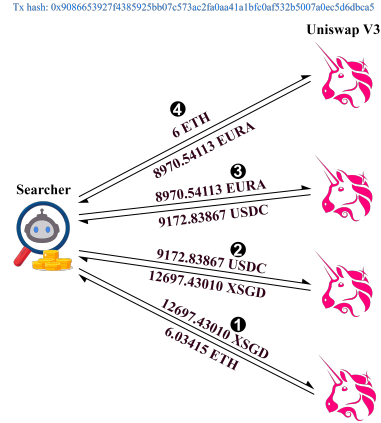
\begin{figure}
    \centering
    \resizebox{0.8\columnwidth}{!}{ 
        \begin{tikzpicture}[every node/.style={align=center, font=\footnotesize}]
        
\node[anchor=west, text=blue, font=\ttfamily\normalsize] at (-9.5, 8) {\href{https://etherscan.io/tx/0x9086653927f4385925bb07c573ac2fa0aa41a1bfc0af532b5007a0ec5d6dbca5}{Tx hash: 0x9086653927f4385925bb07c573ac2fa0aa41a1bfc0af532b5007a0ec5d6dbca5}};

\draw[fill=purple!30, thick] (-7.8, 2.0) rectangle (-5.2, 3.1); 
\node[font=\large] at (-6.5, 2.55) {Searcher};

\draw[fill=white!20, thick] (0.5, 6) rectangle (3.5, 7); 
\node[font=\large] at (2, 6.5) {Uniswap V3};

\draw[fill=white!20, thick] (0.5, 3.8) rectangle (3.5, 4.8); 
\node[font=\large] at (2, 4.3) {Uniswap V3};

\draw[fill=white!20, thick] (0.5, 1) rectangle (3.5, 2); 
\node[font=\large] at (2, 1.5) {Uniswap V3};

\draw[fill=white!20, thick] (0.5, -1) rectangle (3.5, -2); 
\node[font=\large] at (2, -1.5) {Uniswap V3};

\draw[->, thick, >=latex, line width=0.5mm] (-5.2, 3.1) -- (0.5, 6.7); 
\draw[<-, thick, >=latex, line width=0.5mm] (-5.2, 2.9) -- (0.5, 6.5); 

\node[font=\large, rotate=32] at (-1.65, 5.25) {6 ETH\\[0.2cm]8970.54113 EURA};
\node[draw, circle, line width=0.1mm, fill=black, text=white, font=\small] at (-0.8, 6.5) {4}; 

\draw[->, thick, >=latex, line width=0.5mm] (-5.2, 2.8) -- (0.5, 4.4);
\draw[<-, thick, >=latex, line width=0.5mm] (-5.2, 2.6) -- (0.5, 4.2);
\node[font=\large, rotate=16] at (-1.8, 3.65) {8970.54113 EURA\\[0.2cm]9172.83867 USDC};
\node[draw, circle, line width=0.1mm, fill=black, text=white, font=\small] at (-0.2, 4.8) {3};

\draw[<-, thick, >=latex, line width=0.5mm] (0.5, 1.7) -- (-5.2, 2.5);
\draw[->, thick, >=latex, line width=0.5mm] (0.5, 1.5) -- (-5.2, 2.3);
\node[font=\large, rotate=-8] at (-1.8, 1.9) {9172.83867 USDC\\[0.2cm]12697.43010 XSGD};
\node[draw, circle, line width=0.1mm, fill=black, text=white, font=\small] at (-0.1, 2.6) {2}; 

\draw[<-, thick, >=latex, line width=0.5mm] (0.5, -1.4) -- (-5.2, 2.2);
\draw[->, thick, >=latex, line width=0.5mm] (0.5, -1.6) -- (-5.2, 2.0);
\node[font=\large, rotate=-33] at (-2.2, 0.2) {12697.43010 XSGD\\[0.2cm]6.03415 ETH};
\node[draw, circle, line width=0.1mm, fill=black, text=white, font=\small] at (-0.1, 0.1) {1}; 

        \end{tikzpicture}
    }
    \caption{Reverse-ordered cyclic arbitrage transaction.}
    \label{fig:reverse_ordered_cyclic}
\end{figure}

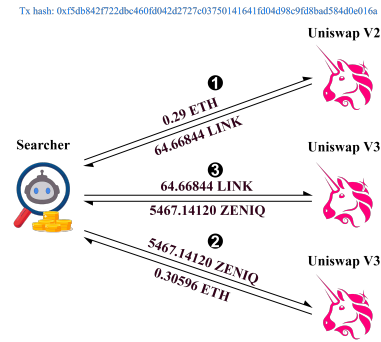
\begin{figure}
    \centering
    \resizebox{0.8\columnwidth}{!}{ 
        \begin{tikzpicture}[every node/.style={align=center, font=\footnotesize}]
        
\node[anchor=west, text=blue, font=\ttfamily\normalsize] at (-9.5, 6) {\href{https://etherscan.io/tx/0xf5db842f722dbc460fd042d2727c03750141641fd04d98c9fd8bad584d0e016a}{Tx hash: 0xf5db842f722dbc460fd042d2727c03750141641fd04d98c9fd8bad584d0e016a}};

\draw[fill=purple!30, thick] (-7.8, 2) rectangle (-5.2, 3); 
\node[font=\large] at (-6.5, 2.5) {Searcher};

\draw[fill=white!20, thick] (0.5, 4) rectangle (3.5, 5); 
\node[font=\large] at (2, 4.5) {Uniswap V2};

\draw[fill=white!20, thick] (0.5, 2) rectangle (3.5, 3); 
\node[font=\large] at (2, 2.5) {Uniswap V3};

\draw[fill=white!20, thick] (0.5, 0) rectangle (3.5, 1); 
\node[font=\large] at (2, 0.5) {Uniswap V3};

\draw[->, thick, >=latex, line width=0.5mm] (-5.2, 2.9) -- (0.5, 4.7); 
\draw[<-, thick, >=latex, line width=0.5mm] (-5.2, 2.7) -- (0.5, 4.5); 
\node[font=\large, rotate=17] at (-1.65, 3.92) {0.29 ETH\\[0.2cm]64.66844 LINK};
\node[draw, circle, line width=0.1mm, fill=black, text=white, font=\small] at (-0.4,4.85) {1}; 

\draw[<-, thick, >=latex, line width=0.5mm] (0.5, 2.6) -- (-5.2, 2.6);
\draw[->, thick, >=latex, line width=0.5mm] (0.5, 2.4) -- (-5.2, 2.4);
\node[font=\large, rotate=0] at (-1.8, 2.5) {64.66844 LINK\\[0.2cm]5467.14120 ZENIQ};
\node[draw, circle, line width=0.1mm, fill=black, text=white, font=\small] at (-0.1, 3.0) {3}; 

\draw[<-, thick, >=latex, line width=0.5mm] (0.5, 0.6) -- (-5.2, 2.3);
\draw[->, thick, >=latex, line width=0.5mm] (0.5, 0.4) -- (-5.2, 2.1);
\node[font=\large, rotate=-15] at (-1.7, 1.12) {5467.14120 ZENIQ\\[0.2cm]0.30596 ETH};
\node[draw, circle, line width=0.1mm, fill=black, text=white, font=\small] at (0.0, 1.6) {2}; 

        \end{tikzpicture}
    }
    \caption{Unordered cyclic arbitrage transaction.}
    \label{fig:unordered_cyclic}
\end{figure}

\begin{figure}
    \centering
    \resizebox{0.8\columnwidth}{!}{ 
        \begin{tikzpicture}[every node/.style={align=center, font=\footnotesize}]
        
\node[anchor=west, text=blue, font=\ttfamily\normalsize] at (-9.5, 4.9) {\href{https://etherscan.io/tx/0x840ecb2b5d55a682afd529138b36e97992eda9706e206237b57ec4697e4f8186}{Tx hash: 0x840ecb2b5d55a682afd529138b36e97992eda9706e206237b57ec4697e4f8186}};

\draw[fill=purple!30, thick] (-7.8, 2.1) rectangle (-5.2, 3.1); 
\node[font=\large] at (-6.5, 2.6) {Searcher};

\draw[fill=white!20, thick] (0.5, 3.2) rectangle (3.5, 4.2); 
\node[font=\large] at (2, 3.7) { Sudoswap};

\draw[fill=white!20, thick] (0.5, 1) rectangle (3.5, 2); 
\node[font=\large] at (2, 1.5) { Sudoswap};

\draw[->, thick, >=latex, line width=0.5mm] (-5.2, 2.9) -- (0.5, 3.8);
\draw[<-, thick, >=latex, line width=0.5mm] (-5.2, 2.7) -- (0.5, 3.6);
\node[font=\large, rotate=8.5] at (-1.8, 3.32) {0.01982 ETH\\[0.2cm]SUDOSQUID [1587]};
\node[draw, circle, line width=0.1mm, fill=black, text=white, font=\small] at (-0.2, 4.2) {1}; 

\draw[<-, thick, >=latex, line width=0.5mm] (0.5, 1.6) -- (-5.2, 2.5);
\draw[->, thick, >=latex, line width=0.5mm] (0.5, 1.4) -- (-5.2, 2.3);
\node[font=\large, rotate=-8] at (-1.8, 1.85) {SUDOSQUID [1587]\\[0.2cm]0.02457 ETH};
\node[draw, circle, line width=0.1mm, fill=black, text=white, font=\small] at (-0.2, 2.45) {2};

        \end{tikzpicture}
    }
    \caption{Cyclic Non-Fungible Token (NFT) arbitrage transaction.}
    \label{fig:cyclic_nft}
\end{figure}
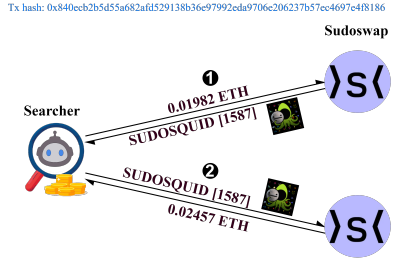

\begin{figure}
    \centering
    \resizebox{0.8\columnwidth}{!}{ 
        \begin{tikzpicture}[every node/.style={align=center, font=\footnotesize}]
        
\node[anchor=west, text=blue, font=\ttfamily\normalsize] at (-8.5,5) {\href{https://etherscan.io/tx/0x98a7e11f4c65e36144e602e52fc267822d476132e781881f4ec3f557c9466b61}{Tx hash: 0x98a7e11f4c65e36144e602e52fc267822d476132e781881f4ec3f557c9466b61}};

\draw[fill=purple!30, thick] (-7,3.2) rectangle (-4.4,4.2); 
\node[font=\large] at (-5.7,3.7) {Searcher 1};

\draw[fill=white!20, thick] (1.2,3.2) rectangle (4.0,4.2); 
\node[font=\large] at (2.6,3.7) { Uniswap v2};

\draw[->, thick, >=latex, line width=0.5mm] (-4.4,3.7) -- (1.2,3.7); 
\node[font=\large] at (-1.7,4.0) {0.13804 ETH}; 

\draw[fill=purple!30, thick] (1.2,-0.5) rectangle (4.0,-1.5); 
\node[font=\large] at (2.6,-1.0) {Searcher 2};

\draw[->, thick, >=latex, line width=0.5mm] (2.5, 3.2) -- (2.5, -0.5); 
\node[font=\large, rotate=-90] at (3.0, 1.3) {156507.03483 \\ COLON};

\draw[fill=white!30, thick] (-7,-0.5) rectangle (-4.4,-1.5); 
\node[font=\large] at (-5.7,-1.0) {Uniswap v3};

\draw[<-, thick, >=latex, line width=0.5mm] (-4.4,-1.0) -- (1.2,-1.0); 
\node[font=\large] at (-1.6,-1.3) {152465.39185 COLON}; 

\draw[<-, thick, >=latex, line width=0.5mm] (-5.7, 3.2) -- (-5.7, -0.5); 
\node[font=\large, rotate=90] at (-6, 1.3) {0.14165 ETH};

        \end{tikzpicture}
    }
    \caption{Cyclic multi-address arbitrage transaction.}
    \label{fig:cyclic_multi_address}
\end{figure}

\subsubsection{Burn and mint}
A burn and mint arbitrage transaction involves the burning of token A and the minting of token B \cite{park2024unraveling}. In this process, a token of interest is swapped for token A, which is then burned, followed by the minting of token B. Token B is then swapped for the token of interest. This type of arbitrage transaction differs from the previously discussed types in two key ways: (1) the token exchange graph does not form a cycle, and (2) burning and minting do not emit events like standard token exchanges. Figure \ref{fig:burn_mint} illustrates an example of a burn and mint arbitrage transaction.

\subsubsection{Set token}
A set token arbitrage transaction involves swapping both set tokens and individual tokens \cite{park2024unraveling}. A set token is composed of multiple underlying tokens. Similar to burn and mint arbitrage, a set token arbitrage transaction does not create a token transfer cycle because redeeming a set token results in more than one token being returned. Figure \ref{fig:set_token} illustrates an example of a set token arbitrage transaction.

\begin{figure}
    \centering
    \resizebox{0.8\columnwidth}{!}{ 
        \begin{tikzpicture}[every node/.style={align=center, font=\footnotesize}]
        
\node[anchor=west, text=blue, font=\ttfamily\normalsize] at (-9.5, 8) {\href{https://etherscan.io/tx/0x08d385d60cbdc2602790a519efda17371b57054828da9a06c522f67f9fa203d4}{Tx hash: 0x08d385d60cbdc2602790a519efda17371b57054828da9a06c522f67f9fa203d4}};

\draw[fill=purple!30, thick] (-7.8, 2.0) rectangle (-5.2, 3.1); 
\node[font=\large] at (-6.5, 2.55) {Searcher};

\draw[fill=white!20, thick] (0.5, 6) rectangle (3.5, 7); 
\node[font=\large] at (2, 6.5) {Shibaswap};

\draw[fill=white!20, thick] (0.5, 3.8) rectangle (3.5, 4.8); 
\node[font=\large] at (2, 4.3) {Null Address};

\draw[fill=white!20, thick] (0.5, 1) rectangle (3.5, 2); 
\node[font=\large] at (2, 1.5) {xSHIB Token};

\draw[fill=white!20, thick] (0.5, -1) rectangle (3.5, -2); 
\node[font=\large] at (2, -1.5) {Shibaswap};

\draw[->, thick, >=latex, line width=0.5mm] (-5.2, 3.1) -- (0.5, 6.7); 
\draw[<-, thick, >=latex, line width=0.5mm] (-5.2, 2.9) -- (0.5, 6.5); 
\node[font=\large, rotate=32] at (-2.3, 4.8) {0.511976 WETH\\[0.2cm]62945271.74928 xSHIB};
\node[draw, circle, line width=0.1mm, fill=black, text=white, font=\small] at (-0.8, 6.5) {1}; 

\draw[->, thick, >=latex, line width=0.5mm] (-5.2, 2.6) -- (0.5, 4.2);
\node[font=\large, rotate=16] at (-1.8, 3.5) {\\[0.2cm]62945271.74928 xSHIB};
\node[draw, circle, line width=0.1mm, fill=black, text=white, font=\small] at (-0.2, 4.8) {2};

\draw[->, thick, >=latex, line width=0.5mm] (0.5, 1.5) -- (-5.2, 2.3);
\node[font=\large, rotate=-8] at (-1.8, 2.1) {62955742.00054 SHIB};
\node[draw, circle, line width=0.1mm, fill=black, text=white, font=\small] at (-0.1, 2.6) {3};

\draw[<-, thick, >=latex, line width=0.5mm] (0.5, -1.4) -- (-5.2, 2.2);
\draw[->, thick, >=latex, line width=0.5mm] (0.5, -1.6) -- (-5.2, 2.0);
\node[font=\large, rotate=-33] at (-2.2, 0.2) {62955742.00054 SHIB\\[0.2cm]0.51660 WETH};
\node[draw, circle, line width=0.1mm, fill=black, text=white, font=\small] at (-0.1, 0.1) {4}; 

        \end{tikzpicture}
    }
    \caption{Burn and mint arbitrage transaction.}
    \label{fig:burn_mint}
\end{figure}

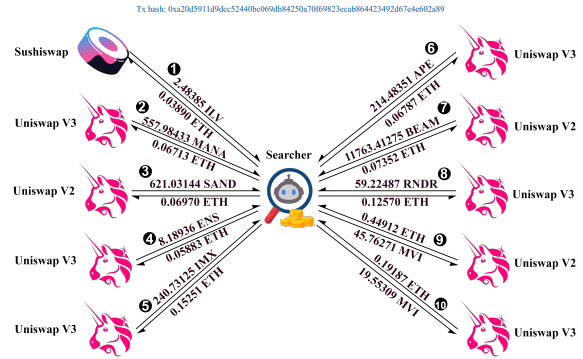
\begin{figure}
    \centering
    \resizebox{\columnwidth}{!}{ 
        \begin{tikzpicture}[every node/.style={align=center, font=\footnotesize}]
        
\node[anchor=west, text=blue, font=\ttfamily\normalsize] at (-13.6, 8) {\href{https://etherscan.io/tx/0xa20d5911d9dcc52440bc069db84250a70f69823ecab864423492d67e4e602a89}{Tx hash: 0xa20d5911d9dcc52440bc069db84250a70f69823ecab864423492d67e4e602a89}};

\draw[fill=purple!30, thick] (-7.8, 1.8) rectangle (-5.2, 3.1); 
\node[font=\large] at (-6.5, 2.45) {Searcher};

\draw[fill=white!20, thick] (0.5, 6.5) rectangle (3.5, 7.5); 
\node[font=\large] at (2, 7) {Uniswap V3};

\draw[fill=white!20, thick] (0.5, 4.5) rectangle (3.5, 5.5); 
\node[font=\large] at (2, 5) {Uniswap V2};

\draw[fill=white!20, thick] (0.5, 2.5) rectangle (3.5, 3.5); 
\node[font=\large] at (2, 3) {Uniswap V3};

\draw[fill=white!20, thick] (0.5, 1.5) rectangle (3.5, 0.5); 
\node[font=\large] at (2, 1) {Uniswap V2};

\draw[fill=white!20, thick] (0.5, -1.5) rectangle (3.5, -0.5); 
\node[font=\large] at (2, -1) {Uniswap V3};

\draw[<-, thick, >=latex, line width=0.5mm] (-5.2, 3.1) -- (0.5, 7.2); 
\draw[->, thick, >=latex, line width=0.5mm] (-5.2, 2.9) -- (0.5, 7.0); 
\node[font=\large, rotate=35] at (-1.5,5.7) {214.48351 APE\\[0.2cm]0.06787 ETH};
\node[draw, circle, line width=0.1mm, fill=black, text=white, font=\small] at (-0.1, 7.2) {6}; 

\draw[<-, thick, >=latex, line width=0.5mm] (-5.2, 2.8) -- (0.5, 5.2);
\draw[->, thick, >=latex, line width=0.5mm] (-5.2, 2.6) -- (0.5, 5.0);
\node[font=\large, rotate=22] at (-1.4, 4.3) {11763.41275 BEAM\\[0.2cm]0.07352 ETH};
\node[draw, circle, line width=0.1mm, fill=black, text=white, font=\small] at (0.2, 5.75) {7};

\draw[->, thick, >=latex, line width=0.5mm] (0.5, 3.2) -- (-5.2, 2.6);
\draw[<-, thick, >=latex, line width=0.5mm] (0.5, 3.0) -- (-5.2, 2.4);
\node[font=\large, rotate=8] at (-1.5, 2.9) {59.22487 RNDR\\[0.2cm]0.12570 ETH};
\node[draw, circle, line width=0.1mm, fill=black, text=white, font=\small] at (0.1, 3.8) {8}; 

\draw[->, thick, >=latex, line width=0.5mm] (0.5, 1.2) -- (-5.2, 2.3);
\draw[<-, thick, >=latex, line width=0.5mm] (0.5, 1.0) -- (-5.2, 2.1);
\node[font=\large, rotate=-10] at (-1.8, 1.5) {0.44912 ETH\\[0.2cm]45.76271 MVI};
\node[draw, circle, line width=0.1mm, fill=black, text=white, font=\small] at (0, 1.8) {9}; 

\draw[->, thick, >=latex, line width=0.5mm] (0.5, -1.0) -- (-5.2, 2.05);
\draw[<-, thick, >=latex, line width=0.5mm] (0.5, -1.2) -- (-5.2, 1.85);
\node[font=\large, rotate=-30] at (-1.7, -0.02) {0.19187 ETH\\[0.2cm]19.55309 MVI};
\node[draw, circle, line width=0.1mm, fill=black, text=white, font=\small] at (0, 0.1) {10};


\draw[fill=white!20, thick] (-16.5, 6.5) rectangle (-13.5, 7.5);
\node[font=\large] at (-15, 7) {Sushiswap};

\draw[fill=white!20, thick] (-16.5, 4.5) rectangle (-13.5, 5.5);
\node[font=\large] at (-15, 5) {Uniswap V3};

\draw[fill=white!20, thick] (-16.5, 2.5) rectangle (-13.5, 3.5);
\node[font=\large] at (-15, 3) {Uniswap V2};

\draw[fill=white!20, thick] (-16.5, 0.5) rectangle (-13.5, 1.5);
\node[font=\large] at (-15, 1) {Uniswap V3};

\draw[fill=white!20, thick] (-16.5, -1.5) rectangle (-13.5, -0.5);
\node[font=\large] at (-15, -1) {Uniswap V3};

\draw[<-, thick, >=latex, line width=0.5mm] (-7.8, 3.1) -- (-13.5, 7.2); 
\draw[->, thick, >=latex, line width=0.5mm] (-7.8, 2.9) -- (-13.5, 7.0); 
\node[font=\large, rotate=-35] at (-11.5,5.65) {2.48385 ILV\\[0.2cm]0.03890 ETH};
\node[draw, circle, line width=0.1mm, fill=black, text=white, font=\small] at (-12.9, 7.2) {1};

\draw[<-, thick, >=latex, line width=0.5mm] (-7.8, 2.8) -- (-13.5, 5.2);
\draw[->, thick, >=latex, line width=0.5mm] (-7.8, 2.6) -- (-13.5, 5.0);
\node[font=\large, rotate=-24] at (-11.4, 4.2) {557.98433 MANA\\[0.2cm]0.06713 ETH};
\node[draw, circle, line width=0.1mm, fill=black, text=white, font=\small] at (-13.0, 5.60) {2};

\draw[->, thick, >=latex, line width=0.5mm] (-13.5, 3.2) -- (-7.8, 2.55);
\draw[<-, thick, >=latex, line width=0.5mm] (-13.5, 3.0) -- (-7.8, 2.4);
\node[font=\large, rotate=-8] at (-11.5, 2.9) {621.03144 SAND\\[0.2cm]0.06970 ETH};
\node[draw, circle, line width=0.1mm, fill=black, text=white, font=\small] at (-13.2, 3.8) {3}; 

\draw[->, thick, >=latex, line width=0.5mm] (-13.5, 1.2) -- (-7.8, 2.3);
\draw[<-, thick, >=latex, line width=0.5mm] (-13.5, 1.0) -- (-7.8, 2.1);
\node[font=\large, rotate=10] at (-11.8, 1.4) {8.18936 ENS\\[0.2cm]0.05883 ETH};
\node[draw, circle, line width=0.1mm, fill=black, text=white, font=\small] at (-13.3, 1.9) {4}; 

\draw[->, thick, >=latex, line width=0.5mm] (-13.5, -1.0) -- (-7.8, 2.05);
\draw[<-, thick, >=latex, line width=0.5mm] (-13.5, -1.2) -- (-7.8, 1.85);
\node[font=\large, rotate=30] at (-11.6, -0.1) {240.73125 IMX\\[0.2cm]0.15251 ETH};
\node[draw, circle, line width=0.1mm, fill=black, text=white, font=\small] at (-13.2, 0.1) {5};

        \end{tikzpicture}
    }
    \caption{Set token arbitrage transaction.}
    \label{fig:set_token}
\end{figure}

\subsubsection{Cross-domain}
A cross-domain arbitrage transaction exploits price discrepancies between asset pairs across two or more domains, such as Layer 1, Layer 2, side-chains, and CEXes \cite{obadia2021unity}. When this arbitrage occurs between an on-chain DEX and an off-chain CEX, it is known as DEX-CEX arbitrage. DEX-CEX arbitrage opportunities arise due to price volatility on CEXes, which can create temporary deviations from prices on on-chain DEXes \cite{heimbach2024non}. If cross-domain arbitrage occurs between DEXes across different blockchains (e.g., Ethereum and BNB chain), it is referred to as cross-chain arbitrage \cite{mazor2024empirical}. When this arbitrage occurs between different execution layers of the same blockchain (e.g., Layer 1 and Layer 2) or between Layer 2 solutions or side-chains, it is referred to as cross-layer arbitrage. When cross-domain arbitrage occurs between different rollups on the same base-layer blockchain (e.g., Arbitrum and Optimism), it is referred to as cross-rollup arbitrage \cite{gogol2024cross}. Cross-domain arbitrage transactions are non-atomic since different domains do not share a unified execution environment. This increases the risk for searchers, as not all transactions may execute successfully, potentially affecting profitability.

\subsection{Liquidation}
Lending protocols allow any user to liquidate loans when the collateral value approaches the threshold. The borrower may prevent liquidation by performing a trade that increases the collateral's value. Alternatively, a non-borrower user can initiate a liquidation transaction to repay the debt and acquire the collateral at a lower/discounted price, thereby protecting the lender from potential losses. This type of liquidation transaction is a value-creating MEV transaction. On the other hand, searchers may exploit these liquidation opportunities to extract MEV by performing value-diverting transactions either in one of the following ways: 1) a searcher monitors the mempool and front-runs the transaction of a competing liquidator to capture their profit, 2) a searcher monitors transactions that would create the liquidation opportunity (referred to as forced liquidation) and then back-runs the transaction to acquire the collateral \cite{weintraub2022flash}, 3) a searcher performs a forced liquidation, by front-running a transaction to create a liquidation opportunity and then back-running another transaction to claim the collateral. Furthermore, a collusion of searchers could manipulate market prices to extract MEV from multiple blocks by performing liquidation transactions \cite{mackinga2022twap}. These types of liquidation transactions have a detrimental impact on the borrowers and the market as it increases collateral requirements. Figure \ref{fig:liquidation} illustrates an example of a liquidation transaction. Liquidations are of two types: fixed spread-based and auction-based \cite{weintraub2022flash}

\begin{figure}
    \centering
    \resizebox{0.8\columnwidth}{!}{ 
        \begin{tikzpicture}[every node/.style={align=center, font=\footnotesize}]
        
\node[anchor=west, text=blue, font=\ttfamily\normalsize] at (-8.3, 6) {\href{https://etherscan.io/tx/0x21585b217911dc7772682229b738173e1a137e11567cbad387cac6ccb04d87b6}{Tx hash: 0x21585b217911dc7772682229b738173e1a137e11567cbad387cac6ccb04d87b6}};

\draw[fill=purple!30, thick] (-7.8, 2) rectangle (-5.2, 3);
\node[font=\large] at (-6.5, 2.5) {Searcher};

\draw[fill=white!20, thick] (1.8, 4) rectangle (5.3, 5); 
\node[font=\large] at (3.5, 4.5) {Aave Protocol V2};

\draw[fill=white!20, thick] (1.8, 2) rectangle (5.3, 3); 
\node[font=\large] at (3.5, 2.5) {Aave Protocol V2};

\draw[fill=white!20, thick] (1.8, 0) rectangle (5.3, 1); 
\node[font=\large] at (3.5, 0.5) {Uniswap v3};

\draw[->, thick, >=latex, line width=0.5mm] (-5.2, 2.8) -- (1.8, 4.6); 
\node[font=\large, rotate=14] at (-2.0, 3.8) {Repays 843.16057 USDT};
\node[draw, circle, line width=0.1mm, fill=black, text=white, font=\small] at (1.0,4.8) {1}; 

\draw[->, thick, >=latex, line width=0.5mm] (1.8, 2.5) -- (-5.2, 2.5);
\node[font=\large, rotate=0] at (-1.6, 2.7) {Receives 0.25109 ETH as collateral};
\node[draw, circle, line width=0.1mm, fill=black, text=white, font=\small] at (1.4, 3.2) {2}; 

\draw[<-, thick, >=latex, line width=0.5mm] (1.8, 0.6) -- (-5.2, 2.4);
\draw[->, thick, >=latex, line width=0.5mm] (1.8, 0.4) -- (-5.2, 2.2);
\node[font=\large, rotate=-15] at (-1.7, 1.4) {0.23913 ETH\\[0.2cm]843.16057 USDT};
\node[draw, circle, line width=0.1mm, fill=black, text=white, font=\small] at (1.3, 1.4) {3}; 

        \end{tikzpicture}
    }
    \caption{Liquidation transaction.}
    \label{fig:liquidation}
\end{figure}
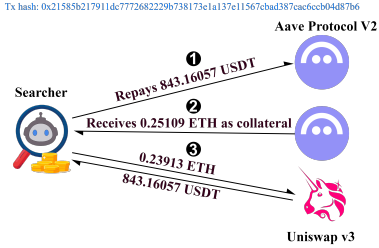

\subsubsection{Fixed spread-based}
Fixed spread-based liquidation operates on a first-come-first-serve basis, where the collateral is awarded to the first user who initiates the liquidation transaction. The liquidator acquires the collateral at a predetermined discount on its market value. However, due to market volatility, the fixed discount may over-compensate or under-compensate the liquidator, leading to an inefficient market.

\subsubsection{Auction-based}
Auction-based liquidation employs a competitive approach where multiple liquidators compete to acquire the collateral by submitting their bids. The collateral is awarded to the user with the highest bid. The auction mechanism mitigates overcompensation risk, as the final price is not based on the market value and is determined through competition. However, it is generally slower than fixed spread-based liquidation. 

\subsection{Time-bandit}
The previously described MEV transactions profit from re-arranging, inserting, or removing transactions in a block that is to be mined. On the contrary, time-bandit MEV transactions aim to extract MEV from already mined block \cite{daian2020flash}. The searcher, on spotting an MEV opportunity in a finalized block, would fork the blockchain by re-mining a block with MEV-acquiring transactions. Time-bandit MEV leads to blockchain instability, as the re-mining of blocks creates competing chains, which can disrupt the finality of transactions and introduce the risk of double-spending. Furthermore, time-bandit can span across domains, enabling a searcher to re-mine blocks in multiple domains, which further destabilizes the network and complicates transaction finality \cite{obadia2021unity}.

Having classified different types of MEV transactions, it is crucial to examine how they can be detected on-chain. Each MEV type exhibits distinct patterns, ordering, and exploitation of victim transactions, which influence the choice of detection approaches. In the following section, we explore various MEV detection approaches and assess their effectiveness in identifying different MEV types.

\section{MEV Detection Approaches}
Detecting MEV is crucial in understanding its potential impact on blockchain security. It enables the evaluation of the effectiveness of MEV mitigation strategies and the development of intelligent frameworks to manage MEV efficiently. Furthermore, quantification of detected MEV allows assessment of financial losses and network instability \cite{qin2022quantifying}. In this section, we discuss various approaches for detecting different types of MEV.

\subsection{Front-running Detection}
Torres et al. \cite{torres2021frontrunner} focused on the detection of displacement and suppression front-running transactions\footnote{\url{https://github.com/christoftorres/Frontrunner-Jones}, accessed on 25 February 2025}. The displacement detection approach begins by identifying the front-running ($T_f$) and victim ($T_v$) transaction pairs using a sliding window that iterates over each transaction within blocks included in the window. Each transaction input bytes is divided into 4-byte n-grams. If 95\%\footnote{100\% matching criteria is not considered as $T_v$ might be encapsulated within $T_f$ along with some metadata.} of n-grams in a transaction corresponds to a previous transaction, a match is reported. For a matched transaction pair, the proposed heuristic then determines if $T_f$ is a displacement transaction if \circled{1} $T_f$ and $T_v$ have different senders as well as receivers (different receivers ensure that bot contracts perform displacement transactions), \circled{2} the gas price of the $T_f$ is larger than that of $T_v$, and \circled{3} ratio between byte sequences of $T_f$ and $T_v$ is at least 25\%. The proposed approach is only limited to displacement transactions performed by bot contracts. Furthermore, it may fail to detect displacement if $T_f$ and $T_v$ do not occur within the same sliding window, leading to false negatives.

The suppression detection approach begins by clustering all transactions with the same receiver within a block, assuming that suppression transactions are directed to the same bot contract \cite{torres2021frontrunner}. The approach then finds clusters that meet the following criteria; \circled{1} more than one transaction exists within a cluster, \circled{2} combined gas consumption of all transactions exceeds 21,000 gas units, and \circled{3} the ratio of gas used to gas limit is over 99\%. Once a cluster is identified, the approach checks for similar clusters in neighboring blocks. The execution trace of the first transaction in the cluster is then analyzed to flag suppression. This approach is limited by the assumption that several suppression transactions should be submitted to suppress a victim transaction. These transactions should span over multiple blocks and must be sent to the same bot contract.

\subsection{Sandwich Detection}
Torres et al. \cite{torres2021frontrunner} proposed a heuristic to detect single DEX and cross-DEX sandwich transactions. The heuristic begins by identifying all possible transfer events\footnote{A transfer event (containing sender's address, receiver's address, and traded amount) is emitted whenever a token is traded.} within a block. The heuristic flags the presence of sandwich transactions if three events (i.e., front-run ($E_f$), victim ($E_v$), and back-run ($E_b$)) satisfy the following conditions: \circled{1} Sender of $E_f$ and $E_v$ is the receiver of $E_b$, and receiver of $E_f$ is the sender of $E_b$, \circled{2} difference between token value $v_f$ of $E_f$ and $v_b$ of $E_b$ is less than or equal to 1\%, \circled{3} token contract addresses for three events are same, \circled{4} transaction hashes for the events are different, \circled{5} transaction index of $E_f$ is smaller than that of $E_v$ which in turn is smaller than that of $E_b$, and \circled{6} gas price of $E_f$ is greater than that of $E_v$ which in turn is greater than or equal to that of $E_b$. The same heuristic is employed by Weintraub et al. \cite{weintraub2022flash} to detect sandwich transactions executed via Flashbots\footnote{\label{weintraub} \url{https://github.com/a-flashbot-in-the-pan}, accessed on 25 February 2025}. However, this heuristic fails to identify sandwich transactions where the difference between $v_f$ and $v_b$ exceeds 1\%\footnote{Front-run transaction index:7, victim transaction index:9, back-run transaction index:10 (Block 20633424)}. 

To detect similar types of sandwich transactions, Qin et al \cite{qin2022quantifying} proposed a heuristic based on the following conditions: \circled{1} front-run ($T_f$), victim ($T_v$), and back-run ($T_b$) transactions are in the same block, and this order, \circled{2} $T_f$ should map to one and only one $T_b$, \circled{3} swap directions of $T_f$ and $T_v$ are the same, and that of $T_b$ is the opposite, \circled{4} either $T_f$ and $T_b$ have the same sender or same receiver smart contract, and \circled{5} difference between token value $v_f$ swapped in $T_f$ and $v_b$ swapped back in $T_b$ is within 10\% bound. However, the heuristic cannot detect sandwich attacks where the difference between $v_f$ and $v_b$ exceeds 10\%\footnote{Front-run transaction index:21, victim transaction index:22, back-run transaction index:23 (Block 20619305)}.

The heuristics employed by \cite{torres2021frontrunner,weintraub2022flash,qin2022quantifying} fail to identify sandwich transactions where the indexes of $T_f$, $T_v$, $T_b$ are not consecutive\footnote{Front-run transaction index:0, victim transaction index:2, back-run transaction index:3 (Block 20619280)} or multiple sandwich transactions occur within a single block. Furthermore, they are not capable of detecting burger, dagwood, and liquidation sandwich transactions. These limitations result in a high rate of false negatives.

To detect burger and dagwood sandwich transactions, Chi et al. \cite{chi2024remeasuring} proposed a heuristic that begins by parsing all events emitted by each transaction within a block and then identifying potential pair of $T_f$ and $T_b$ that include single or multiple $T_v$ between them. A sandwich is flagged if the following conditions are met: \circled{1} Either $T_f$ and $T_v$ have the same sender, or they are sent to the same smart contract, and \circled{2} swap directions of $T_f$ and $T_v$ are the same, and that of $T_b$ is the opposite. To identify dagwood sandwich transactions, the heuristic traverses through possible combinations of $T_f$ and $T_b$ transactions to find profitable combinations. However, the heuristic fails to identify a dagwood sandwich where swap directions of $T_f$ and $T_v$ are opposite, as illustrated in the example provided by \cite{bartoletti2022maximizing}.

To detect liquidation sandwich transactions, Xiong et al. \cite{xiong2023demystifying} proposed a heuristic based on the following conditions: \circled{1} $T_f$ and $T_b$ add and remove liquidity, respectively, within the same block; \circled{2} $T_v$ is a swap transaction in the same pool where the liquidity is both added and removed; \circled{3} $T_f$ and $T_b$ share the same sender, which is different from the sender of $T_v$; and \circled{4} $T_f$, $T_v$, and $T_b$ have consecutive transaction indexes. The authors revealed that the liquidity provided by the searcher should, on average, be 269 times higher than the swap amount performed by the victim, emphasizing the need for a substantial initial fund for the searcher. However, the heuristic can only detect liquidation sandwich transactions.

The previously discussed heuristics for detecting sandwich transactions \cite{torres2021frontrunner, weintraub2022flash, qin2022quantifying, chi2024remeasuring, xiong2023demystifying} rely on decoding smart contracts to identify function calls and transfer events. As a result, these methods are limited to smart contracts with accessible binary files for decoding. To address this gap, Park et al. \cite{park2024unraveling} proposed a heuristic\footnote{\label{park} \url{https://github.com/etelpmoc/arbinet}, accessed on 25 February 2025} for sandwich detection that focuses on token transfer data instead of contract data. The proposed heuristic first identifies the pairs of transactions ($T_f$ and $T_b$) sharing the same recipient address within a single block. Then, the transaction pairs are filtered out if one of the following conditions is met: \circled{1} only one token is transferred in either of the transactions, \circled{2} if profits for all token transfers in $T_f$ are greater than zero, \circled{3} if $T_b$ have all trades with zero profits, \circled{4} if set of token contract address in $T_f$ and $T_b$ are not same, or \circled{5} total profit from $T_f$ and $T_b$ is less than zero. Compared to Flashhbots with a 95.15\% F1-score, the proposed approach achieved an F1-score of 98.83\%. However, since this approach relies on token transfer data, identifying specific function calls becomes difficult, resulting in false positives. Furthermore, the approach is unable to detect burger, dagwood, and liquidation sandwich transactions.

\subsection{Arbitrage Detection}
Daian et al. \cite{daian2020flash} conducted a preliminary quantification of arbitrage in Ethereum based on a manual analysis of atomic arbitrage transactions. Later, Qin et al. \cite{qin2022quantifying} proposed a heuristic to detect arbitrages. According to this heuristic, a transaction is considered arbitrage if the following conditions are met: \circled{1} the transaction contains more than one swap event, and all events are confined within a single transaction, \circled{2} all swap events must form a loop, with the input of one swap being the output of the previous swap, and the input of the first swap being the output of the last swap, and \circled{3} the input value of each swap event should be less than or equal to the output value of the previous event. This heuristic was also employed by Weintraub et al. \cite{weintraub2022flash} to detect arbitrage transactions through Flashbots and Flash Loans\footref{weintraub}. However, the heuristic has several limitations. It fails to identify a cyclic arbitrage transaction if the value of the input token for a swap exceeds the value of the output token from the previous swap \footnote{Transaction index:2 (Block 14628933)}. Furthermore, it cannot detect reverse-ordered cyclic, unordered cyclic, cyclic NFT, or cyclic multi-address arbitrage transactions. Additionally, because it only considers transactions that form a cycle of swap events, it is unable to identify burn and mint or set token arbitrage transactions.

Wang et al. \cite{wang2022cyclic} proposed a method to identify ordered cyclic arbitrage transactions. This approach detects cycles in swap events where the input of a swap is the same as the output of the previous swap. Transactions are considered profitable arbitrages if the product of exchange rates along the cycle exceeds the commission fees paid. However, this heuristic is limited to detecting only ordered cyclic arbitrage transactions.

Chi et al. \cite{chi2024remeasuring} developed a heuristic for detecting unordered cyclic arbitrage transactions. Their method involves extracting input and output tokens and their values from swap events, creating a directed graph to identify cycles. If the transaction includes cyclic swap events and results in a profit, it is classified as arbitrage. Unlike other approaches \cite{qin2022quantifying,weintraub2022flash,wang2022cyclic}, this heuristic does not require the input of each swap to match the output of the previous swap, enabling it to identify unordered and reverse-ordered cyclic arbitrage transactions. However, since it relies solely on token transfer events, it cannot detect cyclic NFT, burn and mint, or set token arbitrage transactions.

Later, Park et al. \cite{park2024unraveling} introduced a heuristic for detecting cyclic NFT arbitrage transactions involving swaps between ERC-20 tokens and ERC-721 NFTs. They also proposed a graph neural network model\footref{park} to identify ordered cyclic, cyclic multi-address, burn and mint, and set token arbitrage transactions. The model uses features related to acquired profit, the number of tokens sent and received, involved addresses, and the number of transfers sent and received to construct a graph. The authors implemented Graph Convolutional Networks (GCN), GraphSAGE, and Graph Attention Networks (GAT). While their approach achieved high F1-scores of 99.54\% and 97.59\% for training and test data, respectively, compared to Flashbots' 68.65\% and 62.85\%, it may suffer from a high false positive rate due to the dynamic nature of the DEX market. Furthermore, a decrease in the performance of test data compared to train data could be due to overfitting and requires further analysis. The approach fails to detect reverse-ordered and unordered cyclic arbitrage transactions.

To detect non-atomic DEX-CEX arbitrage transactions, Heimbach et al. \cite{heimbach2024non} proposed a heuristic\footnote{\scriptsize \url{https://github.com/liobaheimbach/Non-Atomic-Arbitrage-in-Decentralized-Finance}, accessed on 25 February 2025} based on the following conditions: \circled{1} the transaction executes only one swap in a DEX and does not consume more than 400,000 gas, \circled{2} transaction is submitted via a private pool (more details on private pools in Section \ref{subsec_privatepools}), \circled{3} transaction either includes a direct transfer to the block producer or has a priority fee of at least 1 GWei, \circled{4} transaction is either the first swap executed in the desired direction within a liquidity pool or all preceding transaction must have same recipient (to ensure the desired gains before slippage changes the market conditions), and \circled{5} the swap involves two tokens that are also traded on CEXes. However, the heuristic is underlined by the assumptions that non-atomic DEX-CEX arbitrage transactions are not broadcasted to the public mempool to avoid front-running, and searchers pay high-priority fees or direct transfers to the block producers to ensure the inclusion of their transaction in the desired block. Furthermore, the heuristic only focuses on non-atomic transactions.

\subsection{Liquidation Detection}
Weintraub et al. \cite{weintraub2022flash} developed a method to identify liquidation transactions\footref{weintraub} in lending and borrowing protocols via Flashbots and Flash loans by searching for LiquidationCall and LiquidateBorrow events on the Aave and Compound lending platforms, respectively. Furthermore, Qin et al. \cite{qin2022quantifying} extended the scope by detecting liquidation events across Aave, Compound, and dYdX platforms. However, these approaches are limited to specific platforms.

\subsection{Multi-type MEV Detection}
Varun et al. \cite{varun2022mitigating} proposed a neural network-based pre-chain transaction screening approach to detect front-running and sandwich transactions before they are appended to the blockchain. Their approach specifically targets displacement and suppression front-running and single and cross-DEX sandwich transactions. Their approach involves extracting features for a transaction based on prior transactions and inputting them into a pre-trained Multi-Layer Perceptron (MLP) model to identify MEV transactions. The features extracted include the transaction’s gas price, the mean and standard deviation of transactions' gas prices over the last 10 blocks, the mean and standard deviation of the same account’s historical gas prices, gas token usage, and a predicted gas price. The predicted gas price is obtained using a Long Short-Term Memory (LSTM) model based on the gas prices of the previous 12 non-MEV transactions. Using the MEV transactions data identified by Torres et al. \cite{torres2021frontrunner} for training and validation, the approach achieved accuracies of 85.53\% for suppression front-running transactions, 88.58\% for displacement front-running transactions, and 89.08\% for sandwich transactions. However, the proposed learning-based model may be prone to overfitting as it depends on a tailored dataset and might fail to generalize to a wider range of front-running and sandwich transactions.

Li et al. \cite{li2023demystifying} introduced a learning-based approach to detect sandwich, arbitrage, and liquidation MEV transactions. The authors proposed ACTLIFTER, which identifies MEV transactions within Flashbot bundles by mapping them to asset transfer patterns. Additionally, they developed ACTCLUSTER to discover new categories of MEV transactions within the sandwich, arbitrage, and liquidation types. ACTCLUSTER clusters MEV actions identified by ACTLIFTER by converting the bundle into a matrix where each DeFi action is represented as an action block. This matrix is processed using a Convolutional Neural Network (CNN) to extract features, which are then flattened and fed into three fully connected neural network layers. MEV actions are initially labeled as a sandwich, cyclic arbitrage, or liquidation transaction, with these labels being extended after each clustering round using DBSCAN. ACTLIFTER achieved better detection results compared to Etherscan\footnote{\url{https://etherscan.io/}, accessed on 05 September 2024} and DeFiRanger \cite{wu2021defiranger}. However, ACTLIFTER is limited to identifying swap activities involving a token pair and fails to detect token chains.

Table \ref{tab:detection} presents the capabilities of the proposed detection approaches in detecting different types of MEV transactions. In the context of front-running MEV transactions, no approach focuses on replacement transactions. Regarding sandwich MEV transactions, most works focus on detecting single and cross-DEX sandwich transactions \cite{torres2021frontrunner,weintraub2022flash,qin2022quantifying,chi2024remeasuring,park2024unraveling,varun2022mitigating,li2023demystifying}. Only a few focus on detection of burger \cite{chi2024remeasuring,li2023demystifying}, dagwood \cite{chi2024remeasuring}, and liquidation sandwich \cite{xiong2023demystifying,li2023demystifying} transactions. No single approach is capable of detecting all types of sandwich transactions. Furthermore, all these approaches are limited to single-block sandwich transactions. Consequently, the detected sandwich transactions represent only lower bounds and could result in a high number of false negatives \cite{torres2021frontrunner,qin2022quantifying}. For arbitrage detection, most works focus on ordered cyclic transactions \cite{weintraub2022flash,qin2022quantifying,chi2024remeasuring,park2024unraveling,wang2022cyclic,li2023demystifying}. Additionally, most of the works \cite{weintraub2022flash,qin2022quantifying,chi2024remeasuring,park2024unraveling,li2023demystifying} are limited to atomic arbitrage transactions, where multiple swap events occur within a single transaction. Only \cite{wang2022cyclic} addresses arbitrages performed through both atomic and non-atomic transactions. Lastly, \cite{weintraub2022flash,qin2022quantifying,li2023demystifying} focus on the detection of liquidation MEV transactions.

Many detection approaches can be combined to improve MEV transaction identification. Integrating the rules from heuristic-based detection approaches \cite{torres2021frontrunner, weintraub2022flash, qin2022quantifying, chi2024remeasuring, xiong2023demystifying, park2024unraveling, wang2022cyclic}, enables the detection of multiple types of MEV transactions. However, the combined heuristic-based approach, without introducing additional rules, fails to detect replacement, back-running, multi-block sandwich, multi-address arbitrage, burn and mint arbitrage, set token arbitrage, and time bandit MEV transactions. Similarly, while combining learning-based detection approaches \cite{varun2022mitigating, li2023demystifying, park2024unraveling} enhances MEV detection capabilities, it still cannot detect replacement, back-running, dagwood, multi-block sandwich, unordered cyclic arbitrage, reverse ordered cyclic arbitrage and time bandit MEV transactions. In addition, integrating learning-based approaches is challenging due to differences in feature extraction and detection methodologies.

To summarize, no combined approach is capable of detecting all types of MEV transactions. Furthermore, the detection results presented by the studies discussed above \cite{torres2021frontrunner,weintraub2022flash,qin2022quantifying,chi2024remeasuring,xiong2023demystifying,park2024unraveling,daian2020flash,wang2022cyclic,varun2022mitigating,li2023demystifying} are only estimates of MEV, as there is no definitive ground truth for MEV \cite{qin2022quantifying}. In addition, the works employing heuristic methods \cite{torres2021frontrunner,weintraub2022flash,qin2022quantifying,chi2024remeasuring,xiong2023demystifying,park2024unraveling,wang2022cyclic} represent reactive MEV detection approaches, focusing on quantifying MEV acquired in the past. Estimating potential or future MEV via a proactive approach is challenging because the MEV obtained by each searcher varies based on the optimization techniques used to maximize MEV and the targeted victim transactions \cite{judmayer2022estimating}.

While MEV detection approaches help identify and analyze MEV transactions, managing MEV exploitation requires more than just identification. MEV mitigation strategies aim to proactively minimize or democratize MEV exploitation. In the following section, we analyze different MEV mitigation strategies and assess their effectiveness in limiting different types of MEV transactions.

\begin{table*}[htbp]
\centering
\caption{Capabilities of MEV transactions detection approaches.}
\label{tab:detection}
\scalebox{0.63}{
\renewcommand{\arraystretch}{1.2}
\begin{tabular}{cccccccccccccccccccc}
\hline
\rowcolor[HTML]{EFEFEF} 
\multicolumn{1}{|c|}{\cellcolor[HTML]{EFEFEF}} & \multicolumn{19}{c|}{\cellcolor[HTML]{EFEFEF}MEV transactions} \\ \cline{2-20} 
\rowcolor[HTML]{EFEFEF} 
\multicolumn{1}{|c|}{\cellcolor[HTML]{EFEFEF}} & \multicolumn{3}{c|}{\cellcolor[HTML]{EFEFEF}Front-running} & \multicolumn{1}{c|}{\cellcolor[HTML]{EFEFEF}} & \multicolumn{5}{c|}{\cellcolor[HTML]{EFEFEF}Sandwich} & \multicolumn{8}{c|}{\cellcolor[HTML]{EFEFEF}Arbitrage} & \multicolumn{1}{c|}{\cellcolor[HTML]{EFEFEF}} & \multicolumn{1}{c|}{\cellcolor[HTML]{EFEFEF}} \\ \cline{2-4} \cline{6-18}
\rowcolor[HTML]{EFEFEF} 
\multicolumn{1}{|c|}{\multirow{-3}{*}{\cellcolor[HTML]{EFEFEF}\shortstack{Detection\\ approach}}} & \multicolumn{1}{c|}{\cellcolor[HTML]{EFEFEF}\shortstack{Displa-\\cement}} & \multicolumn{1}{c|}{\cellcolor[HTML]{EFEFEF}\shortstack{Repla-\\cement}} & \multicolumn{1}{c|}{\cellcolor[HTML]{EFEFEF}\shortstack{Suppre-\\ssion}} & \multicolumn{1}{c|}{\multirow{-2}{*}{\cellcolor[HTML]{EFEFEF}\shortstack{Back-\\running}}} & \multicolumn{1}{c|}{\cellcolor[HTML]{EFEFEF}\shortstack{Single\\ DEX}} & \multicolumn{1}{c|}{\cellcolor[HTML]{EFEFEF}\shortstack{Cross-\\DEX}} & \multicolumn{1}{c|}{\cellcolor[HTML]{EFEFEF}Burger} & \multicolumn{1}{c|}{\cellcolor[HTML]{EFEFEF}Dagwood} & \multicolumn{1}{c|}{\cellcolor[HTML]{EFEFEF}\shortstack{Liquid-\\ation}} & \multicolumn{1}{c|}{\cellcolor[HTML]{EFEFEF}\shortstack{Ordered\\ cyclic}} & \multicolumn{1}{c|}{\cellcolor[HTML]{EFEFEF}\shortstack{Reverse-\\ordered\\ cyclic}} & \multicolumn{1}{c|}{\cellcolor[HTML]{EFEFEF}\shortstack{Unordered\\ cyclic}} & \multicolumn{1}{c|}{\cellcolor[HTML]{EFEFEF}\shortstack{Cyclic\\ NFT}} & \multicolumn{1}{c|}{\cellcolor[HTML]{EFEFEF}\shortstack{Cyclic\\ multi-\\address}} & \multicolumn{1}{c|}{\cellcolor[HTML]{EFEFEF}\shortstack{Burn\\ and\\ mint}} & \multicolumn{1}{c|}{\cellcolor[HTML]{EFEFEF}\shortstack{Set\\ token}} & \multicolumn{1}{c|}{\cellcolor[HTML]{EFEFEF}\shortstack{Cross-\\domain}} & \multicolumn{1}{c|}{\multirow{-2}{*}{\cellcolor[HTML]{EFEFEF}\shortstack{Liquid-\\ation}}} & \multicolumn{1}{c|}{\multirow{-2}{*}{\cellcolor[HTML]{EFEFEF}\shortstack{Time-\\bandit}}} \\ \hline

\multicolumn{1}{|c|}{\cite{torres2021frontrunner}} & \multicolumn{1}{c|}{$\circlelefthalfblack^1$} & \multicolumn{1}{c|}{$\mdwhtcircle$} & \multicolumn{1}{c|}{$\circlelefthalfblack^2$} & \multicolumn{1}{c|}{$\mdwhtcircle$} & \multicolumn{1}{c|}{$\circlelefthalfblack^3$} & \multicolumn{1}{c|}{$\circlelefthalfblack^3$} & \multicolumn{1}{c|}{$\mdwhtcircle$} & \multicolumn{1}{c|}{$\mdwhtcircle$} & \multicolumn{1}{c|}{$\mdwhtcircle$} & \multicolumn{1}{c|}{$\mdwhtcircle$} & \multicolumn{1}{c|}{$\mdwhtcircle$} & \multicolumn{1}{c|}{$\mdwhtcircle$} & \multicolumn{1}{c|}{$\mdwhtcircle$} & \multicolumn{1}{c|}{$\mdwhtcircle$} & \multicolumn{1}{c|}{$\mdwhtcircle$} & \multicolumn{1}{c|}{$\mdwhtcircle$} & \multicolumn{1}{c|}{$\mdwhtcircle$} & \multicolumn{1}{c|}{$\mdwhtcircle$} & \multicolumn{1}{c|}{$\mdwhtcircle$} \\ \hline

\multicolumn{1}{|c|}{\cite{weintraub2022flash}} & \multicolumn{1}{c|}{$\mdwhtcircle$} & \multicolumn{1}{c|}{$\mdwhtcircle$} & \multicolumn{1}{c|}{$\mdwhtcircle$} & \multicolumn{1}{c|}{$\mdwhtcircle$} & \multicolumn{1}{c|}{$\circlelefthalfblack^3$} & \multicolumn{1}{c|}{$\circlelefthalfblack^3$} & \multicolumn{1}{c|}{$\mdwhtcircle$} & \multicolumn{1}{c|}{$\mdwhtcircle$} & \multicolumn{1}{c|}{$\mdwhtcircle$} & \multicolumn{1}{c|}{$\circlelefthalfblack^4$} & \multicolumn{1}{c|}{$\mdwhtcircle$} & \multicolumn{1}{c|}{$\mdwhtcircle$} & \multicolumn{1}{c|}{$\mdwhtcircle$} & \multicolumn{1}{c|}{$\mdwhtcircle$} & \multicolumn{1}{c|}{$\mdwhtcircle$} & \multicolumn{1}{c|}{$\mdwhtcircle$} & \multicolumn{1}{c|}{$\mdwhtcircle$} & \multicolumn{1}{c|}{$\mdblkcircle$} & \multicolumn{1}{c|}{$\mdwhtcircle$} \\ \hline

\multicolumn{1}{|c|}{\cite{qin2022quantifying}} & \multicolumn{1}{c|}{$\mdwhtcircle$} & \multicolumn{1}{c|}{$\mdwhtcircle$} & \multicolumn{1}{c|}{$\mdwhtcircle$} & \multicolumn{1}{c|}{$\mdwhtcircle$} & \multicolumn{1}{c|}{$\circlelefthalfblack^5$} & \multicolumn{1}{c|}{$\circlelefthalfblack^5$} & \multicolumn{1}{c|}{$\mdwhtcircle$} & \multicolumn{1}{c|}{$\mdwhtcircle$} & \multicolumn{1}{c|}{$\mdwhtcircle$} & \multicolumn{1}{c|}{$\circlelefthalfblack^4$} & \multicolumn{1}{c|}{$\mdwhtcircle$} & \multicolumn{1}{c|}{$\mdwhtcircle$} & \multicolumn{1}{c|}{$\mdwhtcircle$} & \multicolumn{1}{c|}{$\mdwhtcircle$} & \multicolumn{1}{c|}{$\mdwhtcircle$} & \multicolumn{1}{c|}{$\mdwhtcircle$} & \multicolumn{1}{c|}{$\mdwhtcircle$} & \multicolumn{1}{c|}{$\mdblkcircle$} & \multicolumn{1}{c|}{$\mdwhtcircle$} \\ \hline

\multicolumn{1}{|c|}{\cite{chi2024remeasuring}} & \multicolumn{1}{c|}{$\mdwhtcircle$} & \multicolumn{1}{c|}{$\mdwhtcircle$} & \multicolumn{1}{c|}{$\mdwhtcircle$} & \multicolumn{1}{c|}{$\mdwhtcircle$} & \multicolumn{1}{c|}{$\mdblkcircle$} & \multicolumn{1}{c|}{$\mdblkcircle$} & \multicolumn{1}{c|}{$\mdblkcircle$} & \multicolumn{1}{c|}{$\circlelefthalfblack^6$} & \multicolumn{1}{c|}{$\mdwhtcircle$} & \multicolumn{1}{c|}{$\mdblkcircle$} & \multicolumn{1}{c|}{$\mdblkcircle$} & \multicolumn{1}{c|}{$\mdblkcircle$} & \multicolumn{1}{c|}{$\mdwhtcircle$} & \multicolumn{1}{c|}{$\mdwhtcircle$} & \multicolumn{1}{c|}{$\mdwhtcircle$} & \multicolumn{1}{c|}{$\mdwhtcircle$} & \multicolumn{1}{c|}{$\mdwhtcircle$} & \multicolumn{1}{c|}{$\mdwhtcircle$} & \multicolumn{1}{c|}{$\mdwhtcircle$} \\ \hline

\multicolumn{1}{|c|}{\cite{xiong2023demystifying}} & \multicolumn{1}{c|}{$\mdwhtcircle$} & \multicolumn{1}{c|}{$\mdwhtcircle$} & \multicolumn{1}{c|}{$\mdwhtcircle$} & \multicolumn{1}{c|}{$\mdwhtcircle$} & \multicolumn{1}{c|}{$\mdwhtcircle$} & \multicolumn{1}{c|}{$\mdwhtcircle$} & \multicolumn{1}{c|}{$\mdwhtcircle$} & \multicolumn{1}{c|}{$\mdwhtcircle$} & \multicolumn{1}{c|}{$\mdblkcircle$} & \multicolumn{1}{c|}{$\mdwhtcircle$} & \multicolumn{1}{c|}{$\mdwhtcircle$} & \multicolumn{1}{c|}{$\mdwhtcircle$} & \multicolumn{1}{c|}{$\mdwhtcircle$} & \multicolumn{1}{c|}{$\mdwhtcircle$} & \multicolumn{1}{c|}{$\mdwhtcircle$} & \multicolumn{1}{c|}{$\mdwhtcircle$} & \multicolumn{1}{c|}{$\mdwhtcircle$} & \multicolumn{1}{c|}{$\mdwhtcircle$} & \multicolumn{1}{c|}{$\mdwhtcircle$} \\ \hline

\multicolumn{1}{|c|}{\cite{park2024unraveling}} & \multicolumn{1}{c|}{$\mdwhtcircle$} & \multicolumn{1}{c|}{$\mdwhtcircle$} & \multicolumn{1}{c|}{$\mdwhtcircle$} & \multicolumn{1}{c|}{$\mdwhtcircle$} & \multicolumn{1}{c|}{$\mdblkcircle$} & \multicolumn{1}{c|}{$\mdblkcircle$} & \multicolumn{1}{c|}{$\mdwhtcircle$} & \multicolumn{1}{c|}{$\mdwhtcircle$} & \multicolumn{1}{c|}{$\mdwhtcircle$} & \multicolumn{1}{c|}{$\mdblkcircle$} & \multicolumn{1}{c|}{$\mdwhtcircle$} & \multicolumn{1}{c|}{$\mdwhtcircle$} & \multicolumn{1}{c|}{$\mdblkcircle$} & \multicolumn{1}{c|}{$\mdblkcircle$} & \multicolumn{1}{c|}{$\mdblkcircle$} & \multicolumn{1}{c|}{$\mdblkcircle$} & \multicolumn{1}{c|}{$\mdwhtcircle$} & \multicolumn{1}{c|}{$\mdwhtcircle$} & \multicolumn{1}{c|}{$\mdwhtcircle$} \\ \hline

\multicolumn{1}{|c|}{\cite{wang2022cyclic}} & \multicolumn{1}{c|}{$\mdwhtcircle$} & \multicolumn{1}{c|}{$\mdwhtcircle$} & \multicolumn{1}{c|}{$\mdwhtcircle$} & \multicolumn{1}{c|}{$\mdwhtcircle$} & \multicolumn{1}{c|}{$\mdwhtcircle$} & \multicolumn{1}{c|}{$\mdwhtcircle$} & \multicolumn{1}{c|}{$\mdwhtcircle$} & \multicolumn{1}{c|}{$\mdwhtcircle$} & \multicolumn{1}{c|}{$\mdwhtcircle$} & \multicolumn{1}{c|}{$\mdblkcircle$} & \multicolumn{1}{c|}{$\mdwhtcircle$} & \multicolumn{1}{c|}{$\mdwhtcircle$} & \multicolumn{1}{c|}{$\mdwhtcircle$} & \multicolumn{1}{c|}{$\mdwhtcircle$} & \multicolumn{1}{c|}{$\mdwhtcircle$} & \multicolumn{1}{c|}{$\mdwhtcircle$} & \multicolumn{1}{c|}{$\mdwhtcircle$} & \multicolumn{1}{c|}{$\mdwhtcircle$} & \multicolumn{1}{c|}{$\mdwhtcircle$} \\ \hline

\multicolumn{1}{|c|}{\cite{varun2022mitigating}} & \multicolumn{1}{c|}{$\mdblkcircle$} & \multicolumn{1}{c|}{$\mdwhtcircle$} & \multicolumn{1}{c|}{$\mdblkcircle$} & \multicolumn{1}{c|}{$\mdwhtcircle$} & \multicolumn{1}{c|}{$\mdblkcircle$} & \multicolumn{1}{c|}{$\mdblkcircle$} & \multicolumn{1}{c|}{$\mdwhtcircle$} & \multicolumn{1}{c|}{$\mdwhtcircle$} & \multicolumn{1}{c|}{$\mdwhtcircle$} & \multicolumn{1}{c|}{$\mdwhtcircle$} & \multicolumn{1}{c|}{$\mdwhtcircle$} & \multicolumn{1}{c|}{$\mdwhtcircle$} & \multicolumn{1}{c|}{$\mdwhtcircle$} & \multicolumn{1}{c|}{$\mdwhtcircle$} & \multicolumn{1}{c|}{$\mdwhtcircle$} & \multicolumn{1}{c|}{$\mdwhtcircle$} & \multicolumn{1}{c|}{$\mdwhtcircle$} & \multicolumn{1}{c|}{$\mdwhtcircle$} & \multicolumn{1}{c|}{$\mdwhtcircle$} \\ \hline

\multicolumn{1}{|c|}{\cite{li2023demystifying}} & \multicolumn{1}{c|}{$\mdwhtcircle$} & \multicolumn{1}{c|}{$\mdwhtcircle$} & \multicolumn{1}{c|}{$\mdwhtcircle$} & \multicolumn{1}{c|}{$\mdwhtcircle$} & \multicolumn{1}{c|}{$\mdblkcircle$} & \multicolumn{1}{c|}{$\mdblkcircle$} & \multicolumn{1}{c|}{$\mdblkcircle$} & \multicolumn{1}{c|}{$\mdwhtcircle$} & \multicolumn{1}{c|}{$\mdblkcircle$} & \multicolumn{1}{c|}{$\mdblkcircle$} & \multicolumn{1}{c|}{$\mdwhtcircle$} & \multicolumn{1}{c|}{$\mdwhtcircle$} & \multicolumn{1}{c|}{$\mdwhtcircle$} & \multicolumn{1}{c|}{$\mdwhtcircle$} & \multicolumn{1}{c|}{$\mdwhtcircle$} & \multicolumn{1}{c|}{$\mdwhtcircle$} & \multicolumn{1}{c|}{$\mdwhtcircle$} & \multicolumn{1}{c|}{$\mdblkcircle$} & \multicolumn{1}{c|}{$\mdwhtcircle$} \\ \hline

\multicolumn{1}{|c|}{\cite{heimbach2024non}} & \multicolumn{1}{c|}{$\mdwhtcircle$} & \multicolumn{1}{c|}{$\mdwhtcircle$} & \multicolumn{1}{c|}{$\mdwhtcircle$} & \multicolumn{1}{c|}{$\mdwhtcircle$} & \multicolumn{1}{c|}{$\mdwhtcircle$} & \multicolumn{1}{c|}{$\mdwhtcircle$} & \multicolumn{1}{c|}{$\mdwhtcircle$} & \multicolumn{1}{c|}{$\mdwhtcircle$} & \multicolumn{1}{c|}{$\mdwhtcircle$} & \multicolumn{1}{c|}{$\mdwhtcircle$} & \multicolumn{1}{c|}{$\mdwhtcircle$} & \multicolumn{1}{c|}{$\mdwhtcircle$} & \multicolumn{1}{c|}{$\mdwhtcircle$} & \multicolumn{1}{c|}{$\mdwhtcircle$} & \multicolumn{1}{c|}{$\mdwhtcircle$} & \multicolumn{1}{c|}{$\mdwhtcircle$} & 
\multicolumn{1}{c|}{$\circlelefthalfblack^7$} & 
\multicolumn{1}{c|}{$\mdwhtcircle$} & \multicolumn{1}{c|}{$\mdwhtcircle$} \\ \hline

\multicolumn{20}{l}{{\parbox{1.48\linewidth}{$\mdblkcircle$: detected; $\mdwhtcircle$: not detected; $\circlelefthalfblack$: partially detected; $^1$if $T_f$ and $T_v$ occur within the same block range window; $^2$if multiple suppression transactions are sent to the same bot contract; $^3$if difference between $v_f$ and $v_b$ is $\leq$ 1\%; $^4$if the input value of a swap event $\leq$ the output value of previous event; $^5$if difference between $v_f$ and $v_b$ is $\leq$ 10\%; $^6$if $T_f$ and $T_v$ has the same swap directions; $^7$only DEX-CEX}}}

\end{tabular}
}
\end{table*}

\section{MEV Mitigation Strategies}
To mitigate the negative impacts of MEV, whether by reducing its occurrence or democratizing its exploitation, various strategies have been developed. These strategies focus on designing fair transaction ordering policies, executing privacy-preserving transactions, or developing protocols for processing transactions using private pools. In this section, we explore different categories of MEV mitigation strategies, evaluating their effectiveness in addressing different types of MEV while maintaining decentralization and security.

\subsection{Transaction Ordering} \label{subsec_ordering}
Searchers extract MEV by manipulating the ordering of their transactions in a block with respect to victim transactions. This ordering is influenced by the incentives, such as priority fees and private transfers, offered to block producers. By offering greater incentives, searchers can prioritize their transactions and increase their profits. To promote fairness and mitigate MEV, various transaction ordering strategies have been proposed. These strategies are based on transaction submission or arrival times \cite{raikwar2023fairness,mamageishvili2023buying,gans2023cryptography}, random or deterministic sequencer \cite{sinai2024q,doe2023incentive}, or Market Maker (MM) designs \cite{ciampi2022fairmm,moosavi2022lissy}.

The most well-known strategy based on submission or arrival times is order fairness mechanisms such as receive, send, approximate, and block order fairness \cite{raikwar2023fairness}. Receive order fairness states that if a majority (more than 50\%) of validators receive transaction $T_1$ before $T_2$, then $T_1$ should precede $T_2$ in the final order \cite{kursawe2021wendy}. Send order fairness states that if $T_1$ is sent before $T_2$, then $T_1$ should precede $T_2$ in the final order. Intel SGX Trusted Execution Environment (TEE) \cite{zheng2021survey} or ARM Trustzone \cite{pinto2019demystifying} could be used to verify and trust the timestamps of these transactions \cite{ciampi2024universal}. However, due to network delays, achieving strict send order fairness is often infeasible \cite{ciampi2024universal}. To address this, approximate order fairness relaxes strict sequencing of transactions $T_1$ and $T_2$ if validators receive them within a specified time window. If the majority of validator nodes receive $T_1$ at least $\epsilon$ rounds earlier than $T_2$, then all honest validator nodes should not deliver $T_2$ until $T_1$ has been delivered \cite{kelkar2020order}. These order fairness mechanisms primarily consider front-running, back-running, sandwich, arbitrage, and liquidation MEV transactions by focusing on transaction-level ordering. To address time-bandit MEV, block order fairness ensures that if $T_1$ and $T_2$ arrive within a time window and a majority of validators receive $T_1$ first, then $T_1$ should not be placed in a later block than $T_2$ \cite{kursawe2021wendy,kelkar2020order}. Temporal order fairness strategies can be further extended by discretizing the time into fixed periods measured in time units or block numbers \cite{gans2023cryptography}. If multiple searchers submit competing transactions, then disputes can be resolved using the Solomonic approach\footnote{\url{https://github.com/solomonic-mechanism}, accessed on 25 February 2025} \cite{noyes2006judgment}. This approach assumes that each transaction initiator (whether searcher or legitimate user) knows its own status. An initiator (selected randomly) is given the opportunity to withdraw their transaction. If the transaction is withdrawn, the other initiator's transaction is executed, else the transaction amount is burned. A Legitimate initiator incurs a loss if forced to withdraw, while a searcher does not, discouraging illegitimate claims. However, this approach is only suitable for mitigating replacement front-running. Furthermore, implementing and managing the random selection and token-burning mechanisms can be complex, and as the competition among searchers increases, costs and latency also rise.

The order fairness \cite{raikwar2023fairness,gans2023cryptography} strategies rely solely on transaction timestamps, potentially giving an unfair advantage to searchers with low-latency infrastructure \cite{alipanahloo2024maximal}. To address this, Mamageishvili et al. \cite{mamageishvili2023buying} proposed TimeBoost, an ordering strategy based on both transaction timestamps and gas fees. TimeBoost assigns a score for each transaction received within a specific time interval and orders them in descending order. The score is calculated by subtracting the transaction arrival time from a bidding function that considers both bid amounts and total bids in the system. Higher bid increases transaction priority, prompting searchers to compete by resubmitting transactions with progressively increasing bids, which may cause network congestion and instability. Furthermore, TimeBoost could disproportionately favor wealthy searchers who can invest in low-latency infrastructure and afford to place high bids.

To prevent favoring wealthy searchers, transaction ordering strategies based on random \cite{sinai2024q} or deterministic \cite{doe2023incentive} sequencers have been proposed. Sinai and Hoh In \cite{sinai2024q} proposed the Quantum Random Transaction Ordering Protocol (Q-RTOP), which utilizes a quantum random number generator to finalize transaction order. However, Q-RTOP cannot prevent block producers from reordering the randomized transactions, as verifying true randomness remains challenging. On the other hand, Doe et al. \cite{doe2023incentive} proposed a deterministic sequencer strategy for transaction ordering based on transaction confidentiality and delay tolerance. The authors introduced a Weighted Sequencing Service (WSS) that computes the weight for each transaction based on delay tolerance, confidentiality, and transaction workload. Transactions are then ordered based on their weights. However, searchers may manipulate delay tolerance and confidentiality requirements of their transactions to influence the ordering.

In the context of transaction ordering in AMM designs, Ciampi et al. \cite{ciampi2022fairmm} introduced Fair Market Maker (FairMM), an AMM built on their proposed $\Sigma$-Trade protocol. FairMM utilizes off-chain communication for transaction ordering. In $\Sigma$-Trade protocol, a buyer creates a smart contract specifying the required token amount and locks it on the blockchain. The buyer then sends their identity (without token details) to the seller off-chain, and the seller responds with the exchange rate. If the buyer agrees, the contract initiates the trade by forwarding trade details to a MM algorithm. The MM can then accept or decline to trade. If the MM conducts trades with other traders in between, a transaction reordering will be flagged. While FairMM does not explicitly prevent transaction reordering, it discourages malicious activities by publicly flagging any misbehavior. Furthermore, it cannot mitigate back-running transactions since a searcher can execute a transaction that back-runs the victim transaction without requiring any reordering. Empirical analysis shows that, compared to the popular AMM Uniswap, FairMM results in constant gas costs, higher throughput, and faster trade execution. Similarly, Moosavi and Clark \cite{moosavi2022lissy} proposed Lissy\footnote{\url{https://github.com/MadibaGroup/2020-Orderbook}, accessed on 25 February 2025}, an alternative to traditional AMM for fair, efficient, and transparent token trading. Unlike continuous trading in an AMM, Lissy employs a call market mechanism for periodic trading, collecting orders over a set period and executing them simultaneously at a single clearing price, referred to as batch auctions \cite{budish2015high}. This approach eliminates front-running, where traders exploit timing differences to gain an advantage. However, Lissy faces scalability issues in the main Ethereum network due to periodic execution.

\subsection{Privacy-preserving Public Pools}
One strategy to mitigate MEV is restricting transaction visibility, as searchers utilize transaction information from the public mempool to acquire MEV. Consequently, privacy-preserving strategies employ a blind ordering approach, which obscures transaction details from the public mempool before finalizing the transaction order. Once the transaction order is committed, the transaction details are revealed, allowing the transactions to be executed. This mechanism is also known as the commit-and-reveal approach or content-agnostic ordering \cite{raikwar2023fairness}. A well-known privacy-preserving strategy for MEV mitigation is threshold decryption \cite{alipanahloo2024maximal}. In this, a user encrypts the transaction using a global public key and the transaction order is committed to the blockchain. Later, the transaction is decrypted by a committee of decryptors, each holding a share of the decryption key. The transaction details remain private until a sufficient number of committee members, meeting the set threshold, decrypt the transaction. Once threshold decryption occurs, the transactions are executed in the sequence committed on the blockchain.

Kavousi et al. \cite{kavousi2023blindperm} extended the threshold decryption method by proposing BlindPerm, a framework designed to enhance security against MEV transaction reordering. BlindPerm shuffles the order of committed transactions using a seed generated through a permutation technique. This added layer of permutation ensures that even after the transactions are decrypted, their order remains concealed, providing additional protection against MEV exploits. A similar approach is employed by Piet et al. \cite{piet2023mevade}, where encrypted transactions are randomly shuffled using a randomness beacon\footnote{\url{https://github.com/sanjayss34/geth-random-order}, accessed on 25 February 2025}. This beacon utilizes Ethereum's RANDAO algorithm, which generates a random value by combining shares of input from a committee. The input is generally a digital signature of known information. However, some threshold decryption approaches might allow the decryption of pending transactions that are not included in a block. This can expose transaction information, potentially enabling MEV extraction if a searcher is part of the committee of decryptors. Furthermore, threshold decryption introduces high bandwidth overheads because each committee member must propagate their share of the decrypted transaction for every encrypted transaction in the network. This increases the amount of data that needs to be transmitted and processed, leading to network congestion and scalability constraints. Considering an example scenario of 500 transactions, each of size 64 bytes, and a committee of 100 members, the additional data load on the network due to threshold decryption would be lower-bounded at 1632 KB if the threshold is set at 51\%, and upper-bounded at 3200 KB if the threshold is set at 100\%. 

To reduce the bandwidth overheads associated with threshold decryption, Momeni et al. \cite{momeni2022fairblock} proposed FairBlock\footnote{\url{https://github.com/fairblock}, accessed on 25 February 2025}, which uses a global public key and a block identifier (for instance, block index) to encrypt transactions. During the decryption phase, FairBlock generates a shared key for the entire block based on the block identifier, rather than requiring a shared key for each individual transaction as in threshold decryption. This approach significantly reduces bandwidth overheads and improves scalability. Experimental results show that the message size of FairBlock is between 0.4\% and 5.26\% of the message size required by threshold decryption approaches. However, FairBlock introduces a risk of a single point of failure. This is because if the shared key is compromised, all transactions within the blocks could be exposed to MEV exploitation. 

In contrast to encryption-based privacy-preserving MEV mitigation strategies \cite{kavousi2023blindperm,piet2023mevade,momeni2022fairblock}, Kamphuis et al. \cite{kamphuis2023revisiting} proposed Context Preference Robustness (CPR) protocol, which generates transactions within a time-lock puzzle, rather than encrypting transactions. Block producers commit to these transactions and order them based on the time lock, with transaction details revealed after a user-defined time interval. CPR operates on two chains: a control chain and a sanitized chain. The control chain maintains a ledger of time-locked transactions, while the sanitized chain contains the ledger of transactions finalized on the control chain. Transaction details remain hidden until the control chain reaches a depth of k blocks, where k is a user-defined parameter. To prevent block producers from delaying transactions for MEV exploitation, CPR invalidates old transactions, discouraging delays since invalid transactions would not lead to MEV profits. However, legitimate users may lose trust if their transactions are not executed due to such delays. Furthermore, despite the risk of losing MEV, block producers might still attempt to delay transactions, causing inefficiencies. Moreover, the dual-chain structure may introduce complexities and overheads.

\subsection{Private Pools}  \label{subsec_privatepools}
To democratize MEV, private pool mitigation strategies focus on privately submitting transactions to block producers rather than the public mempool. The Flashbots project\footnote{\url{https://www.flashbots.net/}, accessed on 05 September 2024} introduced the concept of private pools and relays. The searchers submit their transactions to relays off-chain, avoiding public mempool exposure. Relays collect these transactions in a private mempool, where the gas fee bids remain sealed, unlike the open-bid process in PGAs. The relays then forward the transactions to block producers. Block producers pack the transactions (both from public mempool and forwarded by the relays) into blocks, taking into account the block size limit. Transactions are ordered by decreasing gas prices within the block. By replacing the open-bid on-chain auction with a sealed-bid off-chain auction, Flashbots effectively mitigate the issues of network congestion and higher gas prices associated with PGAs \cite{mohan2024blockchains}.

The first Flashbots block was mined at block height 11,834,049 on February 11, 2021, after the project launched earlier that year \cite{weintraub2022flash}. Experimental results demonstrated that the sealed-bid auction approach results in lower average bidding prices compared to the conventional open-bid mechanism, contributing to more stable gas prices \cite{jin2023first}. However, the reliance on centralized relays within Flashbots raises trust concerns \cite{capponi2023private}, and there is a lack of accountability regarding potential leakage of builders' private transactions \cite{lyu2022empirical}. Furthermore, Weintraub et al. \cite{weintraub2022flash} argue that the Flashbots project may not optimally democratize and transparently distribute MEV. Their findings suggest that the benefits of Flashbots are disproportionately skewed toward block producers, with searchers sometimes experiencing negative profits, thus failing to achieve fair profit distribution.

After Ethereum's transition from PoW to PoS, an architectural change introduced PBS \cite{mohan2024blockchains, alipanahloo2024maximal}, dividing the block producer role into two distinct roles performed by block builder and block proposer. In PBS, searchers submit transactions privately to block builders. They can also submit a group of ordered transactions, called bundles, that should be mined together in a specific order to ensure they capture MEV. Transactions from users other than searchers can also be submitted privately to the block builder. The block builder creates a block from both public and private transactions, aiming to maximize gas fees. The block is then sent to relays, which act as intermediaries between the block builder and the block proposer. The relay validates the block and forwards only the block header information (excluding transaction details) to the proposer. The block proposer selects the most profitable block (based on block rewards) from those received from multiple relays and returns the signed block to the chosen relay. The proposer then receives the block contents and broadcasts them to the network. Searchers may entice builders to include their bundles in a block by offering higher transaction fees, sharing a portion of the MEV, and/or providing private incentives. In turn, builders might encourage proposers to choose their blocks by offering a share of transaction fees/MEV obtained from searchers and/or by providing other private incentives. However, PBS suffers from the risk of centralization, as a builder relay collusion could indulge in block censoring or MEV exploitation activities \cite{ramos2023mev}. Furthermore, although PBS benefits the participating actors, it has exacerbated the impact of MEV on regular network users. To address this, Babel et al. \cite{babel2024prof} proposed PROF (Protected Order Flow), which introduces a sequencer and bundle merger to mitigate harmful MEV while ensuring profitability for PBS participants. The sequencer collects private transactions from users and orders them into a bundle. The bundle merger, operating at PBS relays, appends this bundle to the most profitable block from builders. The proposer includes the PROF-enriched block for finalization. The authors further introduced PROF-Share, which redistributes MEV profits from back-running back to users. However, in congestion scenarios where the builder-generated bundles occupy the full block capacity, the PROF bundle may be excluded, leading to transaction delays or failures if it does not offer competitive fees to builders and proposers to include the PROF bundle.

Previously discussed mitigation strategies reduce the risk of MEV by transaction ordering. In contrast, intent-centric approaches focus on MEV-resistant transaction generation. For instance, CoWSwap\footnote{\url{https://cow.fi/cow-protocol}, accessed on 14 February 2025} allows users to submit intents rather than submitting detailed transaction information. Third-party entities, referred to as solvers, then compete to execute the intents in batches. However, intent-centric approaches face a trade-off between privacy and efficiency. Similar to the potential collusion between relays and builders in PBS, solver collusion could lead to centralization, posing a risk to the system's fairness.

Building on the discussion of private pools and their impact on transaction transparency, Messias et al. \cite{messias2023dissecting} highlight that the use of private pools undermines transparency, particularly in terms of contention and prioritization. Contention transparency refers to the ability of users to access information about pending transactions and estimate delays for their own transactions, while prioritization transparency ensures that users are aware of the fees associated with pending transactions and can estimate the necessary fees to ensure their transactions are included in a specific block. In private pools, both pending transactions and their associated fees are hidden from users, making the network more opaque and potentially creating opportunities for block producers to overcharge users \cite{messias2023dissecting}.

Table \ref{tab:mitigation} summarizes the effectiveness of mitigation strategies in addressing different types of MEV. Furthermore, it discusses the limitations of each strategy, providing guidance for developers and policy-makers in designing and implementing optimal MEV mitigation strategies. While these strategies focus on reducing MEV exploitation, alternate approaches modify the underlying economic conditions that create MEV opportunities or develop MEV-resistant DeFi ecosystems. For example, Miqado \cite{qin2023mitigating} mitigates liquidation MEV in lending and borrowing protocols by replacing the fixed spread liquidation mechanism with a supporter-based collateral top-up approach. In traditional DeFi lending protocols, searchers exploit liquidation opportunities, but Miqado delays or mitigates this by allowing external supporters to add collateral to prevent liquidation. The supporter can later get repaid with interest by the borrower or claim the position. However, this approach depends on the availability of supporters, which in turn relies on proper incentive mechanisms. Furthermore, formal verification techniques enable the development of MEV-resistant DeFi protocols. Frameworks such as Clockwork Finance \cite{babel2023clockwork} aid in the development of MEV-resistant DeFi protocols by identifying potential security risks and quantifying extractable value. Similarly, KEVM enables symbolic execution for reasoning about transaction ordering and gas costs, ensuring that smart contracts are formally verified against MEV attack vectors \cite{hildenbrandt2018kevm}.

After discussing MEV mitigation strategies, it is crucial to understand how different types of MEV transactions can be simulated, extracted, and optimized in practice. MEV simulation helps model the behavior of various MEV types and assess the effectiveness of detection approaches and mitigation strategies. In addition, MEV extraction and optimization focus on maximizing the extracted value. In the following section, we explore how MEV simulation, extraction, and optimization methods are applied to better understand and manage MEV activities.

\begin{table*}[htbp]
\centering
\caption{Summary of MEV mitigation strategies.}
\label{tab:mitigation}
\scalebox{0.7}{
\begin{tabular}{cccccccc}
\hline
\rowcolor[HTML]{EFEFEF} 
\multicolumn{1}{|c|}{\cellcolor[HTML]{EFEFEF}} & \multicolumn{6}{c|}{\cellcolor[HTML]{EFEFEF}MEV transactions} & \multicolumn{1}{c|}{\cellcolor[HTML]{EFEFEF}} \\ \cline{2-7}
\rowcolor[HTML]{EFEFEF} 
\multicolumn{1}{|c|}{\multirow{-2}{*}{\cellcolor[HTML]{EFEFEF}\begin{tabular}[c]{@{}c@{}}Mitigation\\ strategy\end{tabular}}} & \multicolumn{1}{c|}{\cellcolor[HTML]{EFEFEF}\shortstack{Front-\\running}} & \multicolumn{1}{c|}{\cellcolor[HTML]{EFEFEF}\shortstack{Back-\\running}} & \multicolumn{1}{c|}{\cellcolor[HTML]{EFEFEF}Sandwich} & \multicolumn{1}{c|}{\cellcolor[HTML]{EFEFEF}Arbitrage} & \multicolumn{1}{c|}{\cellcolor[HTML]{EFEFEF}Liquidation} & \multicolumn{1}{c|}{\cellcolor[HTML]{EFEFEF}\shortstack{Time-\\bandit}} & \multicolumn{1}{c|}{\multirow{-2}{*}{\cellcolor[HTML]{EFEFEF}Limitations}} \\ \hline

\rowcolor[HTML]{EFEFEF} 
\multicolumn{8}{|c|}{\cellcolor[HTML]{EFEFEF}Transaction ordering} \\ \hline

\multicolumn{1}{|p{2.8cm}|}{Order fairness \cite{raikwar2023fairness}} & \multicolumn{1}{c|}{$\mdblkcircle$} & \multicolumn{1}{c|}{$\mdblkcircle$} & \multicolumn{1}{c|}{$\mdblkcircle$} & \multicolumn{1}{c|}{$\mdblkcircle$} & \multicolumn{1}{c|}{$\circlelefthalfblack^1$} & \multicolumn{1}{c|}{$\circlelefthalfblack^2$} & \multicolumn{1}{p{12.5cm}|}{Leads to unfair advantage for searchers with low-latency infrastructure} \\ \hline

\multicolumn{1}{|p{2.8cm}|}{Solomonic \cite{gans2023cryptography}} & \multicolumn{1}{c|}{$\circlelefthalfblack^3$} & \multicolumn{1}{c|}{$\mdwhtcircle$} & \multicolumn{1}{c|}{$\mdwhtcircle$} & \multicolumn{1}{c|}{$\mdwhtcircle$} & \multicolumn{1}{c|}{$\mdwhtcircle$} & \multicolumn{1}{c|}{$\mdwhtcircle$} & \multicolumn{1}{p{12.5cm}|}{Complex and leads to higher latency} \\ \hline

\multicolumn{1}{|p{2.8cm}|}{TimeBoost \cite{mamageishvili2023buying}} & \multicolumn{1}{c|}{$\mdblkcircle$} & \multicolumn{1}{c|}{$\mdblkcircle$} & \multicolumn{1}{c|}{$\mdblkcircle$} & \multicolumn{1}{c|}{$\mdblkcircle$} & \multicolumn{1}{c|}{$\circlelefthalfblack^1$} & \multicolumn{1}{c|}{$\mdwhtcircle$} & \multicolumn{1}{p{12.5cm}|}{Favors wealthy searchers and could lead to network congestion and instability} \\ \hline

\multicolumn{1}{|p{2.8cm}|}{Q-RTOP \cite{sinai2024q}} & \multicolumn{1}{c|}{$\mdblkcircle$} & \multicolumn{1}{c|}{$\mdblkcircle$} & \multicolumn{1}{c|}{$\mdblkcircle$} & \multicolumn{1}{c|}{$\mdblkcircle$} & \multicolumn{1}{c|}{$\circlelefthalfblack^1$} & \multicolumn{1}{c|}{$\mdwhtcircle$} & \multicolumn{1}{p{12.5cm}|}{Proving that the randomized sequence is not rearranged by the block producers is difficult} \\ \hline

\multicolumn{1}{|p{2.8cm}|}{WSS \cite{doe2023incentive}} & \multicolumn{1}{c|}{$\mdblkcircle$} & \multicolumn{1}{c|}{$\mdblkcircle$} & \multicolumn{1}{c|}{$\mdblkcircle$} & \multicolumn{1}{c|}{$\mdblkcircle$} & \multicolumn{1}{c|}{$\circlelefthalfblack^1$} & \multicolumn{1}{c|}{$\mdwhtcircle$} & \multicolumn{1}{p{12.5cm}|}{Could increase latency as the transaction rate rises} \\ \hline

\multicolumn{1}{|p{2.8cm}|}{FairMM \cite{ciampi2022fairmm}} & \multicolumn{1}{c|}{$\mdblkcircle$} & \multicolumn{1}{c|}{$\mdwhtcircle$} & \multicolumn{1}{c|}{$\mdwhtcircle$} & \multicolumn{1}{c|}{$\mdblkcircle$} & \multicolumn{1}{c|}{$\circlelefthalfblack^1$} & \multicolumn{1}{c|}{$\mdwhtcircle$} & \multicolumn{1}{p{12.5cm}|}{Prone to trade-order data manipulation due to off-chain communication} \\ \hline

\multicolumn{1}{|p{2.8cm}|}{Lissy \cite{moosavi2022lissy}} & \multicolumn{1}{c|}{$\mdblkcircle$} & \multicolumn{1}{c|}{$\mdblkcircle$} & \multicolumn{1}{c|}{$\mdblkcircle$} & \multicolumn{1}{c|}{$\mdblkcircle$} & \multicolumn{1}{c|}{$\circlelefthalfblack^1$} & \multicolumn{1}{c|}{$\mdwhtcircle$} & \multicolumn{1}{p{12.5cm}|}{Leads to scalability issues and higher latency due to periodic execution} \\ \hline

\rowcolor[HTML]{EFEFEF} 
\multicolumn{8}{|c|}{\cellcolor[HTML]{EFEFEF}Privacy-preserving public pools} \\ \hline

\multicolumn{1}{|p{2.8cm}|}{BlindPerm \cite{kavousi2023blindperm}} & \multicolumn{1}{c|}{$\mdblkcircle$} & \multicolumn{1}{c|}{$\mdblkcircle$} & \multicolumn{1}{c|}{$\mdblkcircle$} & \multicolumn{1}{c|}{$\mdblkcircle$} & \multicolumn{1}{c|}{$\circlelefthalfblack^1$} & \multicolumn{1}{c|}{$\mdwhtcircle$} & \multicolumn{1}{p{12.5cm}|}{Leads to network congestion and restricts scalability due to threshold decryption} \\ \hline

\multicolumn{1}{|p{2.8cm}|}{RANDAO \cite{piet2023mevade}} & \multicolumn{1}{c|}{$\mdblkcircle$} & \multicolumn{1}{c|}{$\mdblkcircle$} & \multicolumn{1}{c|}{$\mdblkcircle$} & \multicolumn{1}{c|}{$\mdblkcircle$} & \multicolumn{1}{c|}{$\circlelefthalfblack^1$} & \multicolumn{1}{c|}{$\mdwhtcircle$} & \multicolumn{1}{p{12.5cm}|}{Causes network congestion and limits scalability due to threshold decryption} \\ \hline

\multicolumn{1}{|p{2.8cm}|}{FairBlock \cite{momeni2022fairblock}} & \multicolumn{1}{c|}{$\mdblkcircle$} & \multicolumn{1}{c|}{$\mdblkcircle$} & \multicolumn{1}{c|}{$\mdblkcircle$} & \multicolumn{1}{c|}{$\mdblkcircle$} & \multicolumn{1}{c|}{$\circlelefthalfblack^1$} & \multicolumn{1}{c|}{$\mdwhtcircle$} & \multicolumn{1}{p{12.5cm}|}{Poses a single point of failure risk as compromising block identifier could expose the details of all transactions within the block} \\ \hline

\multicolumn{1}{|p{2.8cm}|}{CPR \cite{kamphuis2023revisiting}} & \multicolumn{1}{c|}{$\mdblkcircle$} & \multicolumn{1}{c|}{$\mdblkcircle$} & \multicolumn{1}{c|}{$\mdblkcircle$} & \multicolumn{1}{c|}{$\mdblkcircle$} & \multicolumn{1}{c|}{$\circlelefthalfblack^1$} & \multicolumn{1}{c|}{$\mdwhtcircle$} & \multicolumn{1}{p{12.5cm}|}{Leads to user distrust due to invalidated transactions and introduces complexity and overheads due to dual-chain structure} \\ \hline

\rowcolor[HTML]{EFEFEF} 
\multicolumn{8}{|c|}{\cellcolor[HTML]{EFEFEF}Private pools} \\ \hline

\multicolumn{1}{|p{2.8cm}|}{Flashbots} & \multicolumn{1}{c|}{$\mdblkcircle$} & \multicolumn{1}{c|}{$\mdblkcircle$} & \multicolumn{1}{c|}{$\mdblkcircle$} & \multicolumn{1}{c|}{$\mdblkcircle$} & \multicolumn{1}{c|}{$\mdblkcircle$} & \multicolumn{1}{c|}{$\mdwhtcircle$} & \multicolumn{1}{p{12.5cm}|}{Leads to disproportionate MEV distribution and undermines transparency} \\ \hline

\multicolumn{1}{|p{2.8cm}|}{PBS} & \multicolumn{1}{c|}{$\mdblkcircle$} & \multicolumn{1}{c|}{$\mdblkcircle$} & \multicolumn{1}{c|}{$\mdblkcircle$} & \multicolumn{1}{c|}{$\mdblkcircle$} & \multicolumn{1}{c|}{$\mdblkcircle$} & \multicolumn{1}{c|}{$\mdwhtcircle$} & \multicolumn{1}{p{12.5cm}|}{Poses a centralization risk through builder-relay collusion and undermines transparency} \\ \hline

\multicolumn{1}{|p{2.8cm}|}{PROF \cite{babel2024prof}} & \multicolumn{1}{c|}{$\mdblkcircle$} & \multicolumn{1}{c|}{$\mdblkcircle$} & \multicolumn{1}{c|}{$\mdblkcircle$} & \multicolumn{1}{c|}{$\mdblkcircle$} & \multicolumn{1}{c|}{$\mdblkcircle$} & \multicolumn{1}{c|}{$\mdwhtcircle$} & \multicolumn{1}{p{12.5cm}|}{Potential transaction exclusion in congestion scenarios, poses a centralization risk, and undermines transparency} \\ \hline

\multicolumn{1}{|p{2.8cm}|}{CoWSwap} & \multicolumn{1}{c|}{$\mdblkcircle$} & \multicolumn{1}{c|}{$\mdblkcircle$} & \multicolumn{1}{c|}{$\mdblkcircle$} & \multicolumn{1}{c|}{$\mdblkcircle$} & \multicolumn{1}{c|}{$\mdblkcircle$} & \multicolumn{1}{c|}{$\mdwhtcircle$} & \multicolumn{1}{p{12.5cm}|}{Poses a centralization risk and trade-off between efficiency and privacy} \\ \hline

\multicolumn{8}{l}{\parbox{1.4\linewidth}{{$\mdblkcircle$: considered; $\mdwhtcircle$: not considered; $\circlelefthalfblack$: partially considered; $^1$forced liquidation is not considered when a searcher performs an encrypted transaction to create liquidation opportunity and subsequently back-runs another encrypted transaction to acquire collateral; $^2$only block order fairness considers time-bandit MEV, assuming that a searcher or a collusion of searchers does not control more than 51\% of the network; $^3$displacement and suppression front-running transactions are not considered}}}
\end{tabular}
}
\end{table*}

\section{MEV Simulation and Extraction Methods}
Simulating MEV transactions allows researchers and developers to analyze the impact of various MEV types on the stability of consensus mechanisms \cite{babel2023lanturn}. Furthermore, it aids in evaluating the effectiveness of different MEV detection approaches and mitigation strategies. This section examines different methods for simulating MEV transactions in Ethereum and discusses the works on optimizing MEV extraction.

In the context of MEV simulation, Qin et al. \cite{qin2022quantifying} proposed a method to simulate replacement front-running by duplicating a victim transaction, replacing only the sender's address by the searcher's address. The transaction is then executed locally on the highest block. If profitable, the searcher front-runs the victim transaction. A similar approach was developed by Stucke et al. \cite{stucke2022simulation} to simulate displacement front-running transactions\footnote{\url{https://github.com/zakstucke/ethereum-front-running}, accessed on 25 February 2025}. The approach executes the transactions locally on a Ganache fork to compute profitability. The approach also simulates sandwich transactions. This is by identifying victim transactions that perform swaps with a liquidity pool. It then front-runs the victim transaction by performing the same swap and back-runs it a reverse swap. Bogatyy and his team developed a simulator to perform displacement front-running transactions on the Bancor platform\footnote{\url{https://github.com/bogatyy/bancor}, accessed on 05 September 2024}. The simulator monitors the mempool for a victim transaction and executes a displacement transaction on the local Ethereum node. If profitable, the transaction is submitted to the network with a gas price higher than that of the victim via JSON RPC. The Subway bot\footnote{\url{https://github.com/libevm/subway}, accessed on 05 September 2024} simulates sandwich transactions by monitoring the mempool for profitable opportunities. It determines optimal swap values and gas prices for front-running and back-running transactions, then executes them to capture profit.

Regarding MEV extractions, Schwarz-Schilling et al. \cite{schwarz2023time} and {\"O}z et al. \cite{oz2023time} discovered that the bid value submitted by a block builder in PBS increases with bid submission time. This increase is primarily driven by DEX-CEX arbitrage, where the risk associated with arbitrage decreases later in the slot as the DEX and CEX legs can be executed nearly simultaneously. As the time progresses within the block slot, the arbitrage profit becomes more certain, leading to an increase in the bid. Other non-primary factors for the increase might be the expansion of the public mempool and/or the submission of more private transaction bundles to the proposer. Consequently, \cite{schwarz2023time,oz2023time} suggested that an honest but rational block proposer might delay signing a block to optimize MEV. However, empirical analysis in these studies showed that proposers do not engage in such waiting games, likely due to concerns about potential reputation damage associated with self-serving behaviors. To optimize MEV, Babel and Baker \cite{babel2022strategic} proposed MEV-peri, a peer selection approach that selects peers based on the value of transactions relayed by the peers and the latency of the received transactions. Empirical results reveal that peer selection using MEV-peri extracts more MEV compared to random and latency-based peer selection approaches.

Bartoletti et al. \cite{bartoletti2022maximizing} proposed a method to optimize MEV from dagwood sandwich transactions by modeling a transaction as a single-player single-round game. Similarly, Wang et al. \cite{wang2023n} proposed a method to optimize MEV in the case of burger sandwich transactions. The proposed method involves a transaction selecting algorithm and an optimal attack algorithm. The selecting algorithm uses an iterative approach to determine a set of victim transactions that would yield optimal MEV. The optimal attack algorithm employs binary search to output parameters for an optimal front-running transaction based on the set of victim transactions. Babel et al. \cite{babel2023lanturn} introduced Lanturn\footnote{\url{https://github.com/lanturn-defi/lanturn}, accessed on 25 February 2025}, a learning-based framework for simulating MEV optimization strategies. It consists of an optimizer and a simulation environment. The optimizer, inspired by Genetic Algorithms, uses a bi-loop approach to develop a transaction sequence that maximizes profit. The outer loop determines the optimal order of transactions, while the inner loop fine-tunes transaction variables (e.g., token amounts in a swap transaction). The simulation environment evaluates the developed sequence by computing MEV at the current block height, executing transactions in a FIFO (First In, First Out) order on the local node. Experimental results showed that Lanturn extracts more MEV than Flashbots. Zhou et al. \cite{zhou2021just} introduced DEFIPOSER-ARB and DEFIPOSER-SMT to generate cyclic and non-cyclic arbitrage transactions for MEV extraction. DEFIPOSER-ARB identifies arbitrage opportunities across different DEXes by first constructing a graph of the current blockchain state and detecting cyclic transactions with potential profit. A local search then evaluates the profit from these opportunities, and the most profitable transaction is selected for inclusion in the next block. On the other hand, DEFIPOSER-SMT exploits non-cyclic profit-generating transactions by creating a state transition model of the DeFi system. The paths in the transition model represent potential profitable transaction sequences. Heuristics are then employed to prune the search space and execute the most profitable path. Li et al. \cite{li2023unmasking} introduced a Role-Play strategy, wherein a searcher can simultaneously adopt multiple roles, such as trader, lender, borrower, and yield farmer, to optimize MEV exploitation across multiple DeFi applications. By mathematically modeling the interactions between these roles, the authors demonstrate how a searcher can extract significantly more value than traditional single-role exploits.

Table \ref{tab:extraction} compares different MEV extraction methods based on the types of MEV transactions they consider, their inclusion of transaction fees, token transfer fees, wallet balance constraints, and liquidity pool reserves, as well as their evaluation approach, and optimization scope. Transaction fees refer to the gas fees paid by a searcher when executing MEV transactions. Overlooking these fees in the mathematical model for MEV optimization results in an overestimated profit, as the cost of performing transactions is not deducted from the computed gain. Furthermore, while performing transactions on DEXes, a fraction of the trading fee is paid to the liquidity provider. Accounting for this fee is essential to accurately compute MEV gains. It is also crucial to consider the searcher's wallet balance to ensure that transactions can be successfully executed. Assuming an infinite token balance is unrealistic and does not reflect real-world constraints. Similarly, liquidity pool reserves must be taken into account to guarantee that both searcher and victim transactions can be executed successfully. The optimization scope is categorized as global if multiple MEV transactions, whether from the same or different strategies, are executed while accounting for continuous state changes after each MEV transaction. In contrast, local optimization considers only a single MEV strategy in isolation. However, a local optimization approach can still involve multiple transactions and multiple victims within that single strategy.

\begin{table*}[]
\centering
\caption{Summary of MEV extraction and optimization methods.}
\label{tab:extraction}
\scalebox{0.71}{
\begin{tabular}{|c|c|c|c|c|c|c|c|c|c|c|c|c|}
\hline
\multicolumn{1}{|c|}{\cellcolor[HTML]{EFEFEF}} & \multicolumn{6}{|c|}{\cellcolor[HTML]{EFEFEF}\textbf{MEV transactions}} & \multicolumn{1}{|c|}{\cellcolor[HTML]{EFEFEF}} & \multicolumn{1}{|c|}{\cellcolor[HTML]{EFEFEF}} & \multicolumn{1}{|c|}{\cellcolor[HTML]{EFEFEF}} & \multicolumn{1}{|c|}{\cellcolor[HTML]{EFEFEF}} & \multicolumn{1}{|c|}{\cellcolor[HTML]{EFEFEF}} & \multicolumn{1}{|p{2.6cm}|}{\cellcolor[HTML]{EFEFEF}} \\ \cline{2-7} 
\multicolumn{1}{|p{1.5cm}|}{\multirow{-2}{*}{\cellcolor[HTML]{EFEFEF}\begin{tabular}[c]{@{}c@{}}\textbf{Extraction}\\ \textbf{method}\end{tabular}}} & 
\multicolumn{1}{|p{1cm}|}{\cellcolor[HTML]{EFEFEF}\textbf{Front-running}} & 
\multicolumn{1}{|p{1cm}|}{\cellcolor[HTML]{EFEFEF}\textbf{Back-running}} & 
\multicolumn{1}{|p{1.5cm}|}{\cellcolor[HTML]{EFEFEF}\textbf{Sandwich}} & 
\multicolumn{1}{|p{1.5cm}|}{\cellcolor[HTML]{EFEFEF}\textbf{Arbitrage}} & 
\multicolumn{1}{|p{1.3cm}|}{\cellcolor[HTML]{EFEFEF}\textbf{Liquida-tion}} & 
\multicolumn{1}{|p{1cm}|}{\cellcolor[HTML]{EFEFEF}\textbf{Time-bandit}} & 
\multicolumn{1}{|p{1.6cm}|}{\multirow{-2}{*}{\cellcolor[HTML]{EFEFEF}\begin{tabular}[c]{@{}c@{}}\textbf{Transaction}\\ \textbf{fee}\end{tabular}}} & 
\multicolumn{1}{|p{1.2cm}|}{\multirow{-2}{*}{\cellcolor[HTML]{EFEFEF}\begin{tabular}[c]{@{}c@{}}\textbf{Token}\\ \textbf{transfer}\\ \textbf{fee}\end{tabular}}} & 
\multicolumn{1}{|p{1.5cm}|}{\multirow{-2}{*}{\cellcolor[HTML]{EFEFEF}\begin{tabular}[c]{@{}c@{}}\textbf{Bound on}\\ \textbf{wallet}\\\textbf{balance}\end{tabular}}} & 
\multicolumn{1}{|p{2cm}|}{\multirow{-2}{*}{\cellcolor[HTML]{EFEFEF}\begin{tabular}[c]{@{}c@{}}\textbf{Bound on}\\ \textbf{liquidity}\\\textbf{pool reserve}\end{tabular}}} & 
\multicolumn{1}{|p{2cm}|}{\multirow{-2}{*}{\cellcolor[HTML]{EFEFEF}\textbf{Optimization}}} & 
\multicolumn{1}{|c|}{\multirow{-2}{*}{\cellcolor[HTML]{EFEFEF}\begin{tabular}[c]{@{}c@{}}\textbf{Evaluation}\end{tabular}}} \\ \hline

\cite{babel2022strategic} & $\mdblkcircle$ & $\mdblkcircle$ & $\mdblkcircle$ & $\mdblkcircle$ & $\mdblkcircle$ & $\mdwhtcircle$ & N/A & N/A & N/A & N/A & Local & \begin{tabular}[c]{@{}c@{}}Simulation of\\ real data\end{tabular} \\ \hline

\cite{bartoletti2022maximizing} & $\mdwhtcircle$ & $\mdwhtcircle$ & $\mdblkcircle$ & $\mdwhtcircle$ & $\mdwhtcircle$ & $\mdwhtcircle$ & $\mdwhtcircle$ & $\mdwhtcircle$ & $\mdwhtcircle$ & $\mdwhtcircle$ & Local & No execution \\ \hline

\cite{wang2023n} & $\mdwhtcircle$ & $\mdwhtcircle$ & $\mdblkcircle$ & $\mdwhtcircle$ & $\mdwhtcircle$ & $\mdwhtcircle$ & $\mdwhtcircle$ & $\mdblkcircle$ & $\mdwhtcircle$ & $\mdblkcircle$ & Global & \begin{tabular}[c]{@{}c@{}} Simulation of \\ real data \end{tabular} \\ \hline

\cite{babel2023lanturn} & $\mdblkcircle$ & $\mdblkcircle$ & $\mdblkcircle$ & $\mdblkcircle$ & $\mdblkcircle$ & $\mdwhtcircle$ & $\mdwhtcircle$ & $\mdwhtcircle$ & $\mdblkcircle$ & $\mdwhtcircle$ & Global & \begin{tabular}[c]{@{}c@{}} Simulation of \\ real smart contract \end{tabular} \\ \hline

\cite{zhou2021just} & $\mdwhtcircle$ & $\mdwhtcircle$ & $\mdwhtcircle$ & $\mdblkcircle$ & $\mdwhtcircle$ & $\mdwhtcircle$ & N/A & $\mdblkcircle$ & $\mdblkcircle$ & N/A & N/A & \begin{tabular}[c]{@{}c@{}} Simulation of \\ simplified real data \end{tabular}\\ \hline

\cite{li2023unmasking} & $\mdblkcircle$ & $\mdblkcircle$ & $\mdblkcircle$ & $\mdblkcircle$ & $\mdblkcircle$ & $\mdwhtcircle$ & $\mdwhtcircle$ & $\mdwhtcircle$ & $\mdwhtcircle$ & $\mdblkcircle$ & N/A & \begin{tabular}[c]{@{}c@{}} Simulation of \\ real data \end{tabular}\\ \hline

\multicolumn{13}{l}{\parbox{0.7\linewidth}{{$\mdblkcircle$: considered; $\mdwhtcircle$: not considered; N/A: Not Applicable}}} \\
\end{tabular}
}
\end{table*}

\section{Challenges and Possible Solutions}
Several challenges hinder the effectiveness and accuracy of existing MEV detection approaches and mitigation strategies, considering the evolving nature of MEV within DeFi. Despite significant efforts to curb and quantify MEV, current approaches face limitations due to inherent assumptions in their designs. This section explores key challenges reducing the effectiveness of current MEV detection approaches and mitigation strategies. Furthermore, it presents potential solutions paving the way for a more secure, fair and democratized DeFi system. 

\begin{itemize}
    \item \textbf{Centralization and latency:} MEV mitigation strategies in private pools predominantly focus on Flashbots, but this has been shown to result in a biased distribution of MEV \cite{weintraub2022flash}, leading to a decentralized system. In contrast, mitigation strategies for public pools typically focus on fair transaction ordering or commit-and-reveal methods. To ensure temporal fairness in transaction ordering, current methods often rely on a trusted execution environment (TEE) developed by third-party providers like Intel, which centralizes control and reduces decentralization in DeFi systems. On the other hand, commit-and-reveal strategies introduce decryption overhead, leading to longer transaction processing times and increased latency. Consequently, there is a need for a mitigation strategy that enhances the security of the blockchain ecosystem without compromising decentralization or introducing significant latency.

    \item \textbf{Cross-domain MEV:} Current detection approaches and mitigation strategies primarily focus on MEV extracted from transactions on the Ethereum Mainnet, a Layer 1 solution. However, the detection of cross-domain MEV, which consists of DEX-CEX, cross-chain, cross-layer, and cross-rollup, remains largely unaddressed. \cite{heimbach2024non} has explored DEX-CEX arbitrage detection, while \cite{mazor2024empirical} focuses on cross-chain arbitrage. Moreover, \cite{sjursen2023towards,gogol2024cross} has provided insights into the manual analysis of cross-layer and cross-rollup arbitrage transactions. There is a pressing need for solutions that can proactively detect and mitigate MEV exploitation across multiple domains. One approach for mitigating cross-rollup MEV is the use of sequencers, which control the ordering of transactions within rollups. Mechanisms like transaction ordering and batch auctions (Section \ref{subsec_ordering}) can prevent sequencers from manipulating transaction ordering for MEV exploitation. However, mitigating cross-chain MEV is more complex. In particular, bridging protocols, which are used to transfer tokens and assets between different blockchains, introduce delays, creating new opportunities for cross-chain MEV exploitation. Consequently, it becomes crucial to develop secure and efficient cross-domain communication mechanisms and bridging protocols for mitigating MEV associated with cross-chain, cross-layer, and cross-rollup.

    \item \textbf{Mutli-address MEV:} Most MEV detection approaches assume that front-running and back-running transactions in a sandwich originate from the same sender addresses. This assumption restricts the detection of MEV conducted by a searcher using different addresses or through collusion of searchers. Park et al. \cite{park2024unraveling} addressed this issue for arbitrage MEV transactions. However, it remains unconsidered for sandwich and liquidation transactions. Detection methods should integrate this to minimize false negatives.
    
    \item \textbf{MEV quantification discrepancies:} Various heuristics have been proposed to detect and quantify different types of MEV transactions, such as front-running, sandwich, arbitrage, and liquidation. However, the quantification results for a particular MEV transaction type vary across studies due to differences in heuristic rules, block ranges, and the transaction attributes used. As a result, it is challenging to obtain an accurate estimate of the total MEV acquired. A comprehensive evaluation of these heuristics in a unified setup is needed to better analyze their detection capabilities and improve MEV quantification.

    \item \textbf{Comprehensive MEV detection:} Current MEV detection approaches fail to capture all transaction categories within each MEV type, as shown in Table \ref{tab:detection}. Even when these detection approaches are integrated, some MEV transactions remain undetected due to transaction structural complexity and the rigid heuristic rules of these approaches. Developing a comprehensive detection method is challenging, especially as searchers continuously innovate in the evolving DeFi landscape. Consequently, there is a pressing need to develop a more generalized detection approach that minimizes false negatives. This can be achieved by critically analyzing MEV transaction patterns and detecting MEV based on profits rather than solely on swap patterns.
    
    \item \textbf{Value-creating versus value-diverting MEV:} MEV transactions can either contribute positively to the stability of the DeFi system (value-creating) or cause financial losses and disrupt consensus mechanisms (value-diverting). Current MEV detection and mitigation strategies often overlook this critical distinction. Mitigating value-creating MEV may unknowingly destabilize the DeFi system while detecting such transactions could lead to false positives and inaccurate quantification. Researchers and developers should account for this differentiation when designing MEV detection and mitigation approaches to avoid unintended consequences.

    \item \textbf{Proactive MEV detection:} Most existing MEV detection approaches identify MEV transactions only after they are included in the blockchain (i.e., reactive). Proactive detection, such as pre-chain transaction screening approaches  \cite{varun2022mitigating,li2023demystifying}, aim to enhance the security and fairness of the DeFi ecosystem by identifying MEV transactions before they impact the network. These approaches function as early-warning systems that flag MEV transactions, enabling proactive countermeasures. However, several challenges remain in proactive detection. One major issue is feature extraction complexity, as distinguishing MEV transactions from non-MEV requires identifying dynamic and evolving MEV patterns, particularly in differentiating value-creating from value-diverting MEV transactions. Furthermore, models trained on historical data may quickly become obsolete due to the rapid evolution of MEV strategies, necessitating adaptive learning approaches. Proactive detection may also face scalability challenges, as real-time analysis requires significant computation resources, especially when integrating learning approaches. Lastly, submission of transactions via privacy-preserving public pools or private pools enables MEV transactions to bypass detection approaches, reducing the effectiveness of pre-chain screening.

    \item \textbf{Simulation environment:} MEV extraction and optimization methods are often evaluated in simulated environments or local forked chains, such as Ganache. However, ensuring that these environments accurately represent real-world conditions, such as network latency, price slippage, and mempool behavior, remains a challenge. Financial gains observed in simulations may not align with those in actual networks, and, in the worst case, a searcher could incur losses in real-world scenarios. For instance, a victim transaction that has already been included in a block might still appear in the mempool to a searcher due to network latency. The searcher's MEV transactions based on this victim might not yield the desired financial gains. Consequently, it is critical to model network latency and slippage in MEV simulations. Furthermore, simulating adversarial conditions, where multiple searchers compete to extract and optimize the same MEV opportunities, is essential for more accurate quantification of real-world gains.
    
\end{itemize}

\section{Conclusion and Future Work}
MEV has misaligned the DeFi ecosystem's security, efficiency, and decentralization objectives. While efforts have been made to detect and mitigate MEV, they are often tailored to specific types of MEV transactions. Furthermore, the effectiveness of these detection approaches and mitigation strategies remains uncertain, highlighting the need for a closer examination of their capabilities. Existing surveys on MEV tend to focus only on transaction categorization and mitigation strategies. In addition, these aspects are treated in isolation, leading to a fragmented and incomplete understanding of MEV.

In contrast, we performed a comprehensive survey offering a holistic view of the evolving MEV landscape within the DeFi ecosystem. We introduced a novel taxonomy of MEV transactions that distinguishes between value-creating and value-diverting transactions. Through a critical comparative analysis, we assessed the capabilities of MEV detection approaches in identifying the categorized transaction types, offering insights for developing more generalized detection approaches. Furthermore, through an in-depth analysis of MEV mitigation strategies across different transaction types, we highlighted their limitations and emphasized the need for more effective solutions. As MEV continues to pose risks like financial losses and consensus instability, our exploration of simulation frameworks and extraction techniques provided valuable guidance for modeling MEV in real-world scenarios.

Our findings revealed that challenges such as centralization risks, multi-address MEV, accurate detection, and vulnerabilities in layer-2 solutions remain unresolved despite significant advancements. Addressing these challenges is crucial to fostering a more secure, decentralized, and efficient DeFi environment. Further research is still needed to refine detection approaches, enhance mitigation strategies, and explore new solutions to democratize MEV. Additionally, proactive detection approaches are required to identify MEV transactions in real-time and prevent their execution. This comprehensive survey will guide researchers, developers, policymakers, and stakeholders in analyzing and understanding the MEV landscape, offering insights to curb the negative impacts of MEV while unlocking its potential benefits for the DeFi ecosystem.

While our survey provides a qualitative analysis of various MEV detection approaches and mitigation strategies, future work will focus on incorporating quantitative evaluations for a more detailed and objective assessment. This will help guide practitioners and policymakers in selecting the most effective strategies based on specific use cases.

\bibliographystyle{IEEEtran}
\bibliography{references_revision}

\end{document}